\input harvmac
\overfullrule=0pt
\def\N{{\cal N}}
\def\Qtilde{\tilde Q}
\def\M{{\cal M}}
\def\Vbar{\bar V}
\def\Mbar{\bar{\M}}
\def\mubar{\bar{\mu}}
\def\Nbar{\bar{\N}}
\def\nubar{\bar{\nu}}
\def\homoa{\langle {\cal A}\rangle }
\def\barhomoa{\bar{\langle {\cal A}\rangle} }
\def\dmuphys{d\mu^{(k)}_{\rm phys}}
\def\dmuk{d\mu^{(k)}}
\def\VolUk{{\rm Vol}\big(U(k)\big)}
\def\trN{\tr^{}_{\sst N}\,}
\def\Abar{\bar{A}}
\def\zbar{\bar{z}}
\def\qvac{\langle q \rangle}
\def\qbvac{\langle \bar{q} \rangle}
\def\qtvac{\langle \tilde{q}\rangle}
\def\qtbvac{\langle \bar{\tilde{q}} \rangle}
\def\dmuphystilde{d\tilde\mu^{(k)}_{\rm phys}}

\def\sst{\scriptscriptstyle}

\def\Xsd{X^{\sst\rm SD}}

\def\Xasd{X^{\sst\rm ASD}}

\def\F{{\cal F}_{\sst\rm SW}}

\def\A{{\cal A}}
\def\susy{supersymmetry}
\def\sigmabar{\bar\sigma}

\def\cl{{\,\rm cl}}
\def\lambdabar{\bar\lambda}

\def\psibar{\bar\psi}
\def\sqrtwo{\sqrt{2}\,}
\def\etabar{\bar\eta}

\def\Qbar{\bar Q}
\def\susic{supersymmetric}
\def\vhiggs{{\rm v}}

\def\vbarhiggs{\bar{\rm v}}
\def\vhiggsbar{\bar{\rm v}}

\def\Abar{A^\dagger}

\def\C{{\cal C}}

\def\zero{{\scriptscriptstyle(0)}}
\def\new{{\scriptscriptstyle\rm new}}

\def\uA{\,\lower 1.2ex\hbox{$\sim$}\mkern-13.5mu A}
\def\uX{\,\lower 1.2ex\hbox{$\sim$}\mkern-13.5mu X}
\def\uD{\,\lower 1.2ex\hbox{$\sim$}\mkern-13.5mu {\rm D}}

\def\uF{\,\lower 1.2ex\hbox{$\sim$}\mkern-13.5mu F}
\def\uW{\,\lower 1.2ex\hbox{$\sim$}\mkern-13.5mu W}
\def\uWbar{\,\lower 1.2ex\hbox{$\sim$}\mkern-13.5mu {\overline W}}
\def\uV{\,\lower 1.2ex\hbox{$\sim$}\mkern-13.5mu V}
\def\uv{\,\lower 1.0ex\hbox{$\scriptstyle\sim$}\mkern-11.0mu v}
\def\uPsi{\,\lower 1.2ex\hbox{$\sim$}\mkern-13.5mu \Psi}
\def\uPhi{\,\lower 1.2ex\hbox{$\sim$}\mkern-13.5mu \Phi}
\def\uchi{\,\lower 1.5ex\hbox{$\sim$}\mkern-13.5mu \chi}
\def\Psibar{\bar\Psi}
\def\uPsibar{\,\lower 1.2ex\hbox{$\sim$}\mkern-13.5mu \Psibar}
\def\upsi{\,\lower 1.5ex\hbox{$\sim$}\mkern-13.5mu \psi}
\def\psibar{\bar\psi}
\def\upsibar{\,\lower 1.5ex\hbox{$\sim$}\mkern-13.5mu \psibar}
\def\upsibarzero{\,\lower 1.5ex\hbox{$\sim$}\mkern-13.5mu \psibar^\zero}
\def\ulambda{\,\lower 1.2ex\hbox{$\sim$}\mkern-13.5mu \lambda}
\def\ulambdabar{\,\lower 1.2ex\hbox{$\sim$}\mkern-13.5mu \lambdabar}
\def\ulambdabarzero{\,\lower 1.2ex\hbox{$\sim$}\mkern-13.5mu \lambdabar^\zero}
\def\ulambdabarnew{\,\lower 1.2ex\hbox{$\sim$}\mkern-13.5mu \lambdabar^\new}
\def\D{{\cal D}}
\def\M{{\cal M}}
\def\N{{\cal N}}
\def\Dslash{\,\,{\raise.15ex\hbox{/}\mkern-12mu \D}}
\def\Dbarslash{\,\,{\raise.15ex\hbox{/}\mkern-12mu {\bar\D}}}
\def\delslash{\,\,{\raise.15ex\hbox{/}\mkern-9mu \partial}}
\def\delbarslash{\,\,{\raise.15ex\hbox{/}\mkern-9mu {\bar\partial}}}

\def\hf{{\textstyle{1\over2}}}
\def\quarter{{\textstyle{1\over4}}}

\def\fourth{\quarter}

\def\xibar{\bar\xi}
\def\uAcl{\,\lower 1.2ex\hbox{$\sim$}\mkern-13.5mu A^{}_{\cl}}
\def\uAbarcl{\,\lower 1.2ex\hbox{$\sim$}\mkern-13.5mu A_{\cl}^\dagger}

\def\Atot{{\cal A}_{\rm tot}}
\def\mubar{{\bar\mu}}
\def\nubar{{\bar\nu}}
\def\abar{\bar a}
\def\Foneinst{{\cal F}_1}
\def\K{{\cal K}}
\def\Kt{\tilde{\cal K}}
\def\Ahyp{{{\cal A}_{\rm hyp}}}
\def\Ahypdag{{{\cal A}^\dagger_{\rm hyp}}}
\def\hyp{{\rm hyp}}

\def\Lambdahyp{{\Lambda_\hyp}}
\def\Lambdahypdag{{\Lambda^\dagger_\hyp}}
\def\bigL{{\bf L}}
\def\Lambdabar{\bar\Lambda}
\def\Lambdatot{{\Lambda_{\rm tot}}}
\def\bigZ{Z\!\!\!Z}

\def\dmuonephys{d\mu^{(1)}_{\rm phys}}

\def\bigR{{\rm I}\!{\rm R}}
\def\muhyp{\mu_{\rm hyp}}

\def\dmuonephys{d\mu^{(1)}_{\rm phys}}
\def\K{{\cal K}}
\def\Kt{\tilde\K}

\def\utilde{\tilde u}

\def\Atot{{\A_{\rm tot}}}

\def\uA{\,\lower 1.2ex\hbox{$\sim$}\mkern-13.5mu A}
\def\bigL{{\bf L}}

\def\zero{{\scriptscriptstyle(0)}}
\font\authorfont=cmcsc10 \ifx\answ\bigans\else scaled\magstep1\fi
\divide\baselineskip by 7
\multiply\baselineskip by 6
\advance\topskip by -1in\relax
\def\prenomat{\hbox{hep\hbox{-}th/9804009}}
\Title{$\prenomat$}{\vbox{\centerline{The Instanton Hunter's Guide}
\vskip2pt
\centerline{to Supersymmetric $SU(N)$ Gauge Theories}}
}
\centerline{\authorfont Valentin V. Khoze}
\bigskip
\centerline{\sl Department of Physics, Centre for Particle Theory, 
University of Durham}
\centerline{\sl Durham DH1$\,$3LE UK $\quad$ \tt valya.khoze@durham.ac.uk}
\bigskip
\centerline{\authorfont Michael P. Mattis}
\bigskip
\centerline{\sl Theoretical Division T-8, Los Alamos National Laboratory}
\centerline{\sl Los Alamos, NM 87545 USA$\quad$ \tt mattis@pion.lanl.gov}
\bigskip
\centerline{\authorfont Matthew J. Slater}
\bigskip
\centerline{\sl Department of Physics, Centre for Particle Theory, 
University of Durham}
\centerline{\sl Durham DH1$\,$3LE UK $\quad$ \tt m.j.slater@durham.ac.uk}
\vskip .3in
\def\hf{{\textstyle{1\over2}}}

\def\quarter{{\textstyle{1\over4}}}
\noindent
We present a  compendium of  results for ADHM multi-instantons
in   $SU(N)$ SUSY gauge theories, followed by
  applications to $\N=2$ \susic\ models. Extending recent
$SU(2)$ work, and treating the $\N=1$ and $\N=2$ cases in parallel,
we construct: (i) the ADHM supermultiplet,
(ii) the multi-instanton action, and (iii) the 
collective coordinate integration measure. Specializing to  $\N=2$,
we then give a closed formula for ${\cal F}_k,$ the $k$-instanton
contribution to the prepotential, as a finite-dimensional collective
coordinate integral. This amounts to a weak-coupling
 solution, in quadratures, of
the low-energy dynamics of $\N=2$ SQCD,
without appeal to  duality. As an application,
 we calculate ${\cal F}_1$ for all $SU(N)$ and 
any number of flavors $N_F\,$;
for $N_F<2N-2$ and $N_F=2N-1$ we
confirm previous instanton calculations and agree with the proposed
hyper-elliptic curve solutions. For $N_F=2N-2$ and $N_F=2N$ with $N>3$
we obtain new results, which in the latter case we do not understand
how to reconcile with the curves.
\vskip .1in
\Date{\bf April 1998 } 
\vfil\break

\lref\Sone{ N. Seiberg, Phys. Lett. B206 (1988) 75. }
\lref\ENR{J. Erlich, A. Naqvi and L. Randall, hep-th/9801108.}
\lref\Yank{O. Aharony and S. Yankielowicz, hep-th/9601011,
Nucl. Phys. B473 (1996) 93.}
\lref\AStwo{P. Argyres and A. Shapere, hep-th/9509175,
Nucl. Phys. B461 (1996) 437.}
\lref\dkmsw{N. Dorey, V. Khoze, M. Mattis, M. Slater and
W. Weir, \it Instantons, Higher-Derivative Terms, and Nonrenormalization
Theorems in Supersymmetric Gauge Theories\rm,
hep-th/9706007, Phys. Lett. B408 (1997) 213.}
\lref\Fucito{F. Fucito and G. Travaglini, 
{\it Instanton calculus and nonperturbative relations
 in $N=2$ supersymmetric gauge theories}, 
hep-th/9605215,
Phys. Rev. D55 (1997) 1099.}
\lref\oldyung{A. Yung, Nucl. Phys. B 344 (1990) 73.}
\lref\ADS{I. Affleck, Nucl. Phys. B191 (1981) 429;
 I. Affleck, M. Dine and N. Seiberg, Nucl. Phys. B241
(1984) 493; Nucl. Phys. B256 (1985) 557.  }
\lref\IS{K. Intriligator and N. Seiberg, \it
Lectures on supersymmetric gauge theories and electric-magnetic duality,
\rm hep-th/9509066, 
Nucl. Phys. Proc. Suppl. 45BC (1996) 1 and
55B (1996) 200.}
\lref\Shifman{M. Shifman, 
\it Non-perturbative dynamics in supersymmetric gauge theories,
\rm hep-th/9704114, 
Prog. Part. Nucl. Phys. 39 (1997) 1.}
\lref\RGorig{S. Weinberg, Phys. Lett. 91B (1980) 51;
L. Hall, Nucl. Phys. B178 (1981) 75.}
\lref\FPone{D. Finnell and P. Pouliot,
{\it Instanton calculations versus exact results in 4 dimensional 
SUSY gauge theories},
Nucl. Phys. B453 (95) 225, hep-th/9503115.}
\lref\detrefsO{E. Corrigan, P. Goddard, H. Osborn and S. Templeton,
Nucl.~Phys.~B159 (1979) 469; H. Osborn, Nucl.~Phys.~B159 (1979) 497;
H. Osborn and G. P. Moody, Nucl.~Phys.~B173 (1980) 422;
I. Jack, Nucl.~Phys.~B174 (1980) 526;
I. Jack and H. Osborn, Nucl.~Phys.~B207 (1982) 474.}
\lref\detrefsL{H. Berg and M. Luscher, Nucl.~Phys.~B160 (1979) 281;
H. Berg and J. Stehr, Nucl.~Phys.~B173 (1980) 422 and B175 (1980) 293.}
\lref\AH{M. Atiyah and N. Hitchin, {\it ``The Geometry and Dynamics
of Magnetic Monopoles''}, Princeton University Press (1988).}
\lref\gm{G. Gibbons and N. Manton, {\rm Nucl. Phys.} {\rm B274} (1986) 183.}
\lref\dkmten{N. Dorey, V.V. Khoze and M.P. Mattis, \it
Supersymmetry and the multi-instanton measure\rm, hep-th/9708036;
N. Dorey, T. Hollowood, V.V. Khoze and M.P. Mattis, \it
Supersymmetry and the multi-instanton measure II.
{}From $N=4$ to $N=0$\rm, hep-th/9709072.}
\lref\dadda{A. D'Adda and P. Di Vecchia, 
{\rm Phys. Lett.} {\rm 73B} (1978) 162.}
\lref\GMO{P. Goddard, P. Mansfield, and H. Osborn,  Phys.~Lett.~\rm98B
\rm (1981) 59.}
\lref\NSVZ{ V. A. Novikov, M. A. Shifman, A. I. Vainshtein and
V. I. Zakharov, Nucl Phys. B229 (1983) 394; Nucl. Phys. B229 (1983)
407; Nucl. Phys. B260 (1985) 157. }
\lref\ADHM{M.  Atiyah, V.  Drinfeld, N.  Hitchin and
Yu.~Manin, Phys. Lett. A65 (1978) 185. }
\lref\Osborn{H. Osborn, Ann. Phys. 135 (1981) 373. }
\lref\Osborntwo{H. Osborn, Nucl. Phys. B140 (1978) 45.}
\lref\CGTone{ E. Corrigan, D. Fairlie, P. Goddard and S. Templeton,
    Nucl. Phys. B140 (1978) 31; 
E. Corrigan, P. Goddard and S. Templeton,
Nucl. Phys. B151 (1979) 93.}
\lref\dkmone{N. Dorey, V.V. Khoze and M.P. Mattis, \it Multi-instanton
calculus in $N=2$ supersymmetric gauge theory\rm, hep-th/9603136,
Phys.~Rev.~D54 (1996) 2921.}
\lref\dkmfive{N. Dorey, V.V. Khoze and M.P. Mattis, \it On $N=2$
supersymmetric QCD with $4$ flavors\rm,  hep-th/9611016, 
Nucl.~Phys.~\rm B492 \rm (1997) 607.}
\lref\dkmsix{N. Dorey, V.V. Khoze and M.P. Mattis, \it On mass-deformed $N=4$
supersymmetric Yang-Mills theory\rm,  hep-th/9612231, 
Phys.~Lett.~B396 (1997) 141.}
\lref\dkmfour{N. Dorey, V.V. Khoze and M.P. Mattis, \it Multi-instanton
calculus in $N=2$ supersymmetric gauge theory.
II. Coupling to matter\rm, hep-th/9607202, Phys.~Rev.~D54 (1996) 7832.}
\lref\tHooft{G. 't Hooft, Phys. Rev. D14 (1976) 3432; ibid.
(E) D18 (1978) 2199.}
\lref\Bernard{ C. Bernard, Phys. Rev. D19 (1979) 3013.    }
\lref\CWS{ N. H. Christ, E. J. Weinberg and N. K. Stanton, Phys.
Rev. D18 (1978) 2013. }
\lref\BCGW{ C. Bernard, N. H. Christ, A. Guth and E. J. Weinberg, Phys.
Rev. D16 (1977) 2967. }
\lref\Gates{J. Gates, Nucl. Phys. B238 (1984) 349.}
\lref\WVP{B. deWit and A. Van Proeyen, Nucl. Phys. B245 (1984) 89.}
\lref\SWone{N. Seiberg and E. Witten, 
{\it Electric-magnetic duality, monopole
condensation, and confinement in $N=2$ supersymmetric Yang-Mills  
theory}, 
Nucl. Phys. B426 (1994) 19, (E) B430 (1994) 485,  hep-th/9407087}
\lref\SWtwo{N. Seiberg and E. Witten, 
{\it Monopoles, duality and chiral symmetry breaking
in $N=2$ supersymmetric QCD}, 
Nucl. Phys. B431 (1994) 484,  hep-th/9408099}
\lref\BPST{A. Belavin, A. Polyakov, A. Schwartz and
Y. Tyupkin, Phys. Lett. 59B (1975) 85.
  }
\lref\Cordes{ 
S. Cordes, Nucl. Phys. B273 (1986) 629.  }
\lref\WB{ J. Wess and J. Bagger, {\it Supersymmetry and supergravity}, 
Princeton University Press, 1992
  }
\lref\gilmore{R. Gilmore, \it Lie Groups, Lie Algebras and Some
of Their Applications\rm, Wiley-Interscience 1974.}
\lref\KLTY{A. Klemm, W. Lerche, S. Theisen and S. Yankielowicz,
hep-th/9411048, Phys. Lett. B344 (1995) 169.}
\lref\AF{P.C. Argyres and A.E. Faraggi, hep-th/9411057,
Phys. Rev. Lett. 73 (1995) 3931.}
\lref\HO{A. Hanany and Y. Oz, hep-th/9505075,
Nucl. Phys. B452 (1995) 283.}
\lref\APS{ P.C. Argyres, M.R. Plesser and A. Shapere,
hep-th/9505100, Phys. Rev. Lett. 75 (1995) 1699.}
\lref\MN{J.A. Minahan and D. Nemeschansky,
hep-th/9507032, Nucl. Phys. B464 (1996) 3.}
\lref\MNII{J.A. Minahan and D. Nemeschansky,
hep-th/9601059, Nucl. Phys. B468 (1996) 72.}
\lref\DKP{E. D'Hoker, I.M. Krichever and D.H. Phong,
hep-th/9609041, Nucl. Phys. B489 (1997) 179.}
\lref\Witten{E. Witten, \it Solutions of Four-Dimensional
Field Theories via M Theory, \rm hep-th/9703166
Nucl. Phys. B500 (1997) 3.}
\lref\IStwo{ K. Ito and N. Sasakura,    
hep-th/9609104,  Mod. Phys. Lett. A12 (1997) 205; and
hep-th/9602073, Phys. Lett. B382 (1996) 95}
\lref\Slater{M.J. Slater, hep-th/9701170,
Phys. Lett. B403 (1997) 57.}
\lref\Matone{M. 
Matone, {\it Instantons and recursion relations in 
$N=2$ SUSY gauge theory},
 Phys. Lett. B357 (1995) 342,    hep-th/9506102.  }
\lref\dkmtwo{N. Dorey, V.V. Khoze and M.P. Mattis, 
\it Multi-instanton check of the relation between the prepotential  
${\cal F}$ 
and the modulus $u$ in $N=2$ SUSY Yang-Mills theory\rm,
hep-th/9606199, Phys.~Lett.~B390 (1997) 205.} 
\lref\Amati{See for instance
D. Amati, K. Konishi, Y. Meurice, G. Rossi and G. Veneziano,
Phys. Rep. 162 (1988) 169.}
\lref\AHSW{H. Aoyama, T. Harano, M. Sato and S.Wada,
hep-th/9607076, Phys. Lett. B388 (1996) 331.}
\def\frac#1#2{{ {#1}\over{#2}}}

\def\inst{{\rm inst}}

\def\trtwo{\tr^{}_2\,}

\def\Ubar{\bar U}
\def\wbar{\bar w}

\def\abar{\bar a}
\def\bbar{\bar b}

\def\Deltabar{\bar\Delta}
\def\dalpha{{\dot\alpha}}
\def\dbeta{{\dot\beta}}

\def\Im{{\rm Im}}
\def\sst{\scriptscriptstyle}

\def\F{{\cal F}}

\def\A{{\cal A}}
\def\susy{supersymmetry}
\def\sigmabar{\bar\sigma}

\def\cl{{\,\rm cl}}
\def\lambdabar{\bar\lambda}

\def\psibar{\bar\psi}
\def\sqrtwo{\sqrt{2}\,}
\def\etabar{\bar\eta}

\def\Qbar{\bar Q}
\def\susic{supersymmetric}

\def\vhiggs{{\rm v}}

\def\vbarhiggs{\bar{\rm v}}
\def\vhiggsbar{\bar{\rm v}}

\def\C{{\cal C}}

\def\new{{\scriptscriptstyle\rm new}}

\def\uX{\,\lower 1.2ex\hbox{$\sim$}\mkern-13.5mu X}
\def\uQ{\,\lower 1.2ex\hbox{$\sim$}\mkern-13.5mu Q}
\def\uQtilde{\,\lower 1.2ex\hbox{$\sim$}\mkern-13.5mu \tilde Q}
\def\uD{\,\lower 1.2ex\hbox{$\sim$}\mkern-13.5mu {\rm D}}

\def\uF{\,\lower 1.2ex\hbox{$\sim$}\mkern-13.5mu F}
\def\uW{\,\lower 1.2ex\hbox{$\sim$}\mkern-13.5mu W}
\def\uWbar{\,\lower 1.2ex\hbox{$\sim$}\mkern-13.5mu {\overline W}}
\def\uPhibar{\,\lower 1.2ex\hbox{$\sim$}\mkern-13.5mu {\overline \Phi}}

\def\uV{\,\lower 1.2ex\hbox{$\sim$}\mkern-13.5mu V}
\def\uv{\,\lower 1.0ex\hbox{$\scriptstyle\sim$}\mkern-11.0mu v}
\def\uPsi{\,\lower 1.2ex\hbox{$\sim$}\mkern-13.5mu \Psi}
\def\uPhi{\,\lower 1.2ex\hbox{$\sim$}\mkern-13.5mu \Phi}
\def\uchi{\,\lower 1.5ex\hbox{$\sim$}\mkern-13.5mu \chi}
\def\uchitilde{\,\lower 1.5ex\hbox{$\sim$}\mkern-13.5mu \tilde\chi}
\def\Psibar{\bar\Psi}
\def\uPsibar{\,\lower 1.2ex\hbox{$\sim$}\mkern-13.5mu \Psibar}
\def\upsi{\,\lower 1.5ex\hbox{$\sim$}\mkern-13.5mu \psi}
\def\uq{\,\lower 1.5ex\hbox{$\sim$}\mkern-13.5mu q}
\def\uqtilde{\,\lower 1.5ex\hbox{$\sim$}\mkern-13.5mu \tilde q}
\def\psibar{\bar\psi}
\def\upsibar{\,\lower 1.5ex\hbox{$\sim$}\mkern-13.5mu \psibar}
\def\upsibarzero{\,\lower 1.5ex\hbox{$\sim$}\mkern-13.5mu \psibar^\zero}
\def\ulambda{\,\lower 1.2ex\hbox{$\sim$}\mkern-13.5mu \lambda}
\def\ulambdabar{\,\lower 1.2ex\hbox{$\sim$}\mkern-13.5mu \lambdabar}
\def\ulambdabarzero{\,\lower 1.2ex\hbox{$\sim$}\mkern-13.5mu \lambdabar^\zero}
\def\ulambdabarnew{\,\lower 1.2ex\hbox{$\sim$}\mkern-13.5mu \lambdabar^\new}
\def\D{{\cal D}}
\def\M{{\cal M}}
\def\N{{\cal N}}
\def\Dslash{\,\,{\raise.15ex\hbox{/}\mkern-12mu \D}}
\def\Dbarslash{\,\,{\raise.15ex\hbox{/}\mkern-12mu {\bar\D}}}
\def\delslash{\,\,{\raise.15ex\hbox{/}\mkern-9mu \partial}}
\def\delbarslash{\,\,{\raise.15ex\hbox{/}\mkern-9mu {\bar\partial}}}

\def\hf{{\textstyle{1\over2}}}
\def\quarter{{\textstyle{1\over4}}}

\def\fourth{\quarter}

\def\xibar{\bar\xi}

\def\uAcl{\,\lower 1.2ex\hbox{$\sim$}\mkern-13.5mu A^{}_{\cl}}
\def\uAbarcl{\,\lower 1.2ex\hbox{$\sim$}\mkern-13.5mu A_{\cl}^\dagger}

\def\taueff{\tau_{\rm eff}}

\def\deltafcns{\hbox{$\delta$-functions}}

\newsec{Introduction}

This paper has two overlapping agendas. On the one hand (Secs.~2-6), it is
 a compendium of useful formulae for multi-instanton
calculus in both $\N=1$ and $\N=2$ \susic\ four-dimensional $SU(N)$ or $U(N)$ 
gauge theories. The reader familiar with our earlier
work (principally Refs.~\refs{\dkmone\dkmfour-\dkmten}), which
was restricted to the gauge group $SU(2),$ will find no surprises here:
the key expressions extend in a natural way from $SU(2)$ to
$SU(N)$ or $U(N)$, with a suitable adaptation of notation.

On the other hand (Secs.~7-9), we also present some concrete applications
of this general formalism to $\N=2$ \susic\ theories. Of particular
interest are the models where the number of flavors $N_F$ equals
twice the number of colors, for which the $\beta$-function vanishes.
As will be discussed in Sec.~9, for  $N\ge4$ with
$N_F=2N,$ it is not obvious how the proposed exact solutions to these
models can be reconciled with explicit (multi-)instanton results, such
as those obtained herein.

In general, the $\N=2$ \susic\ models, with $0\le N_F\le 2N,$ are
important not so much for their phenomenological content (they require
adjoint Higgs bosons), but rather for their solubility: exact 
statements at the quantum level can be made both about the spectrum of
BPS states, and about the leading low-energy dynamics of
the Wilsonian effective action
 at energies below the spontaneous
gauge symmetry breaking scale. Extending the seminal work of Seiberg and
Witten on $SU(2)$ gauge theories \refs{\SWone,\SWtwo}, 
many people have proposed exact low-energy
solutions for models with classical simple and product gauge groups and
a variety of matter representations. These solutions involve postulating
spectral curves from which a physical object, the prepotential $\F(A)$
\refs{\Gates,\WVP} ($A$ being the adjoint Higgs), is constructed. In turn, 
the low-energy
effective Lagrangian is constructed from derivatives of the prepotential.

As shown in Ref.~\Sone,
the prepotential admits a multi-instanton expansion:
\eqn\multiexp{\F\ =\ \F_{1\hbox{-}\rm loop}\ +\
\sum_{k=1}^\infty\,\F_k}
where $\F_k$ denotes the contribution of the $k$-instanton sector. In our
earlier $SU(2)$ work, and in the present paper for $SU(N)$, we take a
``bottom up'' approach to the construction of the prepotential, and examine
the \susic\ multi-instantons directly. One immediate aim of this
multi-instanton program, starting with \FPone,
has been to verify (and in certain interesting cases
\refs{\dkmfour,\IStwo\AHSW\dkmfive\dkmsix-\Slater}, to modify) the
proposed exact solutions. On a more conceptual level,
this program has allowed us
 to visualize the sum \multiexp\ in a new and physical way,
 with $\F_k$ expressed
as a definite integral over the bosonic and fermionic
collective coordinates of the
$k$-instanton configuration
 \dkmten. An eventual goal is to apply the knowledge
learned about multi-instantons through comparison with exact solutions
(e.g., the collective coordinate integration measure \dkmten),
to a variety of other models where exact solutions are not 
known \Amati.

This paper is organized as follows. 
In Sec.~2 we review the  construction of the general self-dual
gauge configurations in $SU(N)$ theory, for any
 topological number $k$. This construction is due originally to Atiyah,
Drinfeld, Hitchin and Manin \refs{\ADHM\CGTone\CWS-\Osborn}, and the resulting
configurations are known as ADHM multi-instantons. Here we follow most
closely the derivation given in \CGTone. 

Sections 3-6 are devoted to 
 merging the ADHM construction with \susy. Throughout
these sections we
treat the  $\N=1$ and $\N=2$ cases in parallel. In Sec.~3,
following Ref.~\dkmone, we  construct the supermultiplets
of classical configurations. The classical adjoint fermions were first
derived in \CGTone. For the $\N=2$ case one also requires the classical
adjoint Higgs
bosons; their construction for $SU(2)$
was one of the key results of \refs{\dkmone,\dkmfour}, and
we show how to extend it to $SU(N)$ and/or $U(N)$ 
(depending on whether the sum of the adjoint VEVs is required to vanish).
In Sec.~4 we explain, following \dkmfour, how the $\N=1$ and $\N=2$
\susy\ algebras may be realized directly on the space of unconstrained
bosonic and fermionic ADHM collective coordinates, prior to the imposition
of the polynomial constraints that they are required to obey.
In Sec.~5, generalizing Refs.~\refs{\dkmone,\dkmfour}, we obtain the
$\N=1$ and $\N=2$ multi-instanton actions for $SU(N)$ or $U(N)$ gauge theory
coupled to $N_F$ flavors of
 quark hypermultiplets. And in Sec.~6, following Ref.~\dkmten, we derive
 the  $\N=1$ and
$\N=2$ collective coordinate integration measures. Throughout these
sections we suppress many of the calculational details;
 the reader can find much more information in
Refs.~\refs{\dkmone\dkmfour-\dkmten}, in the context of $SU(2)$.

Sections 7-9 are devoted to applying this general formalism to the $\N=2$
models. In Sec.~7 we give a closed formula for $\F_k$ 
introduced in Eq.~\multiexp, as a finite dimensional integral over the
bosonic and fermionic collective coordinates of the \susic\ 
\hbox{$k$-instanton}
configuration. This expression is a natural generalization to
$SU(N)$ and/or $U(N)$ of the $SU(2)$ result presented in Ref.~\dkmten.
It amounts to a weak-coupling series solution, in quadratures, of the
low-energy dynamics of $SU(N)$ and/or $U(N)$ $\N=2$ SQCD, without appeal
to duality. 

As an application, we calculate in Sec.~8, and analyze in Sec.~9, the
\hbox{1-instanton}
term $\F_1$ for an arbitrary number $N_F$ of quark flavors.\foot{In the
case of $SU(2)$, with $N_F>0$ massless flavors,
 the contributions of odd numbers of instantons
 to the prepotential
is forced to vanish by a $\bigZ_2$ symmetry \refs{\SWtwo,\dkmfour}.
This $\bigZ_2$ symmetry is related to the pseudoreality of $SU(2)$, and
does not apply to $SU(N)$ with $N>2$. 
Thus, the first nontrivial contributions to $\F(A)$ come at
 the $1$-instanton level.}
 This calculation
was accomplished previously by Ito and Sasakura \IStwo,
who made two simplifying assumptions: (1) they assumed that
the final answer depends
only on the VEVs $\{\vhiggs_1,\cdots,\vhiggs_N\}$ of $A$ and not on the
complex conjugate parameters $\{\vbarhiggs_1,\cdots,\vbarhiggs_N\}$
(a property known as holomorphy); and (2) they only extracted
the terms in the solution that become singular in the limit that
 two of the VEVs approach one another.
Here, by using the collective coordinate
measure derived in Sec.~6,
we are able to drop both assumptions. Consequently, we are able to
\it derive \rm rather than assume holomorphy. Moreover, for the special cases
$N_F=2N-2$ and $N_F=2N,$ we are able to extract the ``regular'' terms
in $\F_1$, namely the terms that are nonsingular for all choices of VEVs.
(Ito and Sasakura were able to do so with an explicit integration
 for the special case of the gauge group
$SU(3)$.)
For all other $N_F$ the regular terms are required to vanish for dimensional
reasons, and we recapture the results of Ito and Sasakura
(Sec.~9.1). This calculation
is the first example of the usefulness of our measure.

 Finally, in Secs.~9.2-9.3,  following \IStwo,
 we compare our 1-instanton results against the exact
solutions of $SU(N)$ $\N=2$ SQCD proposed in
 Refs.~\refs{\KLTY\AF\HO\APS\MN\MNII-\Witten}. 
For $N_F<2N-2$ and $N_F=2N-1$ we agree with the proposed
hyper-elliptic curve solutions. As mentioned above,
the comparison is most interesting in the
special instances $N_F=2N$ for which the $\beta$-function vanishes.
In these models
 the instanton calculation gives information relating the microscopic
and effective theories, that is not otherwise obtainable from the
hyper-elliptic curves, nor from $M$-theory \Witten, using present methods.
In fact, whether these curves for  $N\ge4$  with
$N_F=2N$ are uniquely specified is currently an open question
\MNII. Regardless of the resolution of this question, for these
models,
we do not presently know how the (multi-)instanton calculations
can be reconciled even in principle with these curves.

\newsec{The ADHM Construction of the $U(N)$ Multi-Instanton}

In this section we concern ourselves with pure $U(N)$ or $SU(N)$ gauge theory,
without fermions or scalars.
Gauge fields $v_m$ are anti-Hermitian $N\times N$ matrices
and 
$v_{mn}=\partial_m v_n -\partial_n v_m +[v_m , v_n]$ is the field-strength.
In the case of $SU(N)$, these quantities are required to be traceless.

For the special case of $SU(2),$ the ADHM formalism reviewed here is 
slightly different (and a little more complicated) than that adopted
in our earlier papers, as reviewed in Sec.~6 of Ref.~\dkmone. That formalism
is actually the one for the symplectic groups, and exploits the fact that
$SU(2)\equiv Sp(1).$ 
Of course, the two alternative formalisms must give the same 
predictions for physical quantities. But the choices of ADHM
 collective coordinates used in the two approaches are somewhat different,
and no one to our knowledge has worked out the details of the ``dictionary''
linking them.
A useful comparison of the ADHM construction for the different
classical groups may be found in Refs.~\refs{\CGTone-\Osborn}.

\subsec{Construction of the classical gauge field}

The ADHM multi-instanton is the general solution
of the self-duality equation, 
\eqn\selfdual{v_{mn}\ =\ ^*v_{mn}\ ,}
in the sector of topological number (equivalently, winding or
instanton number) $k$, where
\eqn\tch{
k\ =\ -{1 \over 16 \pi^2} \int d^4 x \ \tr_{\sst N}({v_{mn} \ }^*v^{mn}) \ . }
The ADHM construction of such multi-instantons is discussed in
Refs.~\refs{\ADHM\CGTone\CWS-\Osborn}. Here we follow, with minor
 modifications,  the $U(N)$ formalism of Ref.~\CGTone.

 The basic object in the ADHM construction
is the $(N+2k)\times 2k$ complex-valued matrix 
$\Delta_{\sst [N+2k] \times [2k]}$ which is taken to be linear in the 
space-time variable $x_m\,$:\foot{For 
clarity we will occasionally show matrix sizes explicitly,
e.g. the $U(N)$ gauge field will be denoted
$v^m_{\sst [N] \times [N]}$. To
represent matrix multiplication in this notation
we will underline contracted indices: 
$(AB)_{\sst [a] \times [c]} = \ A_{\sst [a] \times \underline{\sst [b]}} \ 
B_{ \underline{\sst [b]}\times [c]}$. Also we adopt the shorthand
$X_{\,[m}Y_{n]}=X_mY_n-X_nY_m$. While 
instantons are conventionally dealt with in Euclidean space, we
always work in Minkowski space to keep
supersymmetry manifest. 
Euclidean sigma matrices are 
$\sigma^m_{\rm E}=(1,-i\tau^a)_{\alpha \dalpha}$, and 
$\sigmabar^{m}_{\rm E}=(1,i\tau^a)^{\dalpha \alpha}$
are their Hermitian conjugates. Here $\tau^{1,2,3}$ are the standard Pauli
matrices. Hermitian conjugation 
$\sigma^m_{\alpha \dalpha} \ \to \sigmabar^{m \, \dalpha \alpha}$ 
will always be assumed
to have been taken in Euclidean space. 
In Minkowski space we use analytic 
continuation: $\sigma^m =(-1,\tau^a)$ and $\sigmabar^m =(-1,-\tau^a)$.
Our conventions are consistent with Wess and Bagger \WB\
and are fully described  in \dkmone.}
\eqn\dlt{\Delta_{\sst [N+2k] \times [2k]} \ (x) \ \equiv
\ \Delta_{\sst [N+2k] \times [k] \times [2]} \ (x) 
\ =\  a_{\sst [N+2k] \times [k] \times [2]} \ + 
\ b_{\sst [N+2k] \times [k] \times \underline{\sst [2]}} \ 
x_{\sst \underline{[2]} \times [2]} \ .}
Here we have represented the $[2k]$ index set as a product of two index sets
$[k] \times [2]$ and have used a quaternionic representation of $x_m$,
\eqn\quat{  x_{\sst [2] \times [2]}\  =\ x_{\alpha\dalpha}\ =\
 \ x_m \ \sigma^m_{\alpha\dalpha}\ ,}
where $\sigma^m_{\alpha\dalpha}$ are the four spin matrices.
It follows that $\partial_m\Delta = b\sigma_m$.
By counting the number of bosonic and fermionic zero modes,
we will soon verify 
that $k$ in Eq.~\dlt\ is indeed the instanton number while
$N$ is the parameter in the gauge group $U(N)$ or $SU(N)$.
As discussed below, 
the  complex-valued constant matrices $a$ and $b$ in Eq. \dlt\ constitute
a (highly overcomplete) set of $k$-instanton collective coordinates.

For generic $x$, the nullspace of the Hermitian conjugate matrix
 $\bar\Delta(x)$
is $N$-dimensional, as it has $N$ fewer rows than columns. 
The basis vectors for this nullspace can be assembled
into an $(N+2k)\times N$ dimensional  complex-valued matrix $U(x)$, 
\eqn\uan{
\Deltabar_{\sst [2k] \times \underline{\sst [N+2k]}} \ 
U_{\sst \underline{[N+2k]} \times [N]} 
= \ 0 = \
\Ubar_{\sst [N] \times \underline{\sst [N+2k]}} \ 
\Delta_{\sst \underline{[N+2k]} \times [2k]} \ ,}
where $U$ is orthonormalized according to
\eqn\udef{\Ubar_{\sst [N] \times \underline{\sst [N+2k]}} 
\ U_{\underline{\sst [N+2k]} \times {\sst [N]}}\ = \ 1_{\sst [N] \times [N]}\
 .  }
In turn, the classical gauge field $v_m$ is constructed from $U$ as
follows. Note first that
for the special case $k=0,$  the antisymmetric
gauge configuration $v_m$  defined by
\eqn\vdef{v_m{}_{\sst [N] \times [N]}\  = \ 
\Ubar_{\sst [N] \times \underline{\sst [N+2k]}} \ \partial_{m}
\ U_{\underline{\sst [N+2k]} \times {\sst [N]}}  }
is ``pure gauge'' (i.e., it is a gauge transformation of the vacuum),
 so that it automatically
solves the self-duality equation \selfdual\
in the vacuum sector. The ADHM ansatz is that Eq.~\vdef\ continues
to give a solution to Eq.~\selfdual, even for nonzero $k$. As we
shall see, this requires
the additional condition
\eqn\dbd{\Deltabar_{\sst [2] \times[k] \times \underline{\sst [N+2k]}} \
\Delta_{\sst \underline{[N+2k]} \times [k] \times[2]} \ = \ 
1_{\sst [2] \times [2]} \ f^{-1}_{\sst [k]\times [k]}  }
where $f$ is an arbitrary $x$-dependent $k\times k$ dimensional
Hermitian matrix. 

To check the validity of the ADHM ansatz, note
that  Eq.~\dbd\ combined with the nullspace condition \uan\
imply the completeness relation
\eqn\cmpl{\Delta_{\sst [N+2k] \times \underline{\sst [k]}
 \times \underline{\sst [2]} } \ 
f_{\underline{\sst [k]} \times \underline{\sst [k]} }\,
\Deltabar_{\sst \underline{[2]} \times \underline{[k]} \times [N+2k]} = \
1_{\sst [N+2k] \times[N+2k]} \ -\ 
U_{\sst [N+2k] \times \underline{\sst [N]}} \ 
\Ubar_{\sst \underline{[N]} \times [N+2k]}\ .   }
With the above relations together with integrations by parts, the expression
for the field strength $v_{mn}$ may  be massaged as follows:
\eqn\sdu{\eqalign{v_{mn}\ &= \ \partial_{\,[m}v_{n]} + \ v_{\,[m} v_{n]} \ = \ 
\partial_{\,[m}(\Ubar\partial_{n]}U)+
(\Ubar\partial_{\,[m}U)(\Ubar\partial_{n]}U)\ =\ 
\partial_{\,[m}\Ubar(1-U\Ubar)\partial_{n]}U    
\cr&=\ \partial_{\,[m}\Ubar \Delta f \Deltabar\partial_{n]}U \ = \ 
\Ubar\partial_{\,[m}\Delta f \partial_{n]}\Deltabar U\ =\ 
\Ubar b \sigma_{[m}\sigmabar_{n]}f \bar{b} U \ =\
4\Ubar b \sigma_{mn}f\bbar U\ .}}
Self-duality of the field strength\foot{In 
Minkowski space the self-dual (SD) and anti-self-dual (ASD)
components of an antisymmetric tensor $X_{mn}$ are projected out using
$\Xsd_{mn}\ =\ \quarter\big(\eta_{mk}\eta_{nl}-\eta_{ml}
\eta_{nk}+i\epsilon_{mnkl}\big)X^{kl}$ and $\Xasd_{mn}=(\Xsd_{mn})^*$,
where $\epsilon_{0123}=-\epsilon^{0123}=-1$. Also, since
$\sigma^{mn}=\fourth\,\sigma^{[\,m}\sigmabar^{n\,]}$ and
$\sigmabar^{mn}=\fourth\,\sigmabar^{[\,m}\sigma^{n\,]}$ are
self-dual and anti-self-dual, respectively \WB, it follows that
$\sigma^{mn\ \beta}_{\phantom{mn}\alpha}X_{mn}
=\sigma^{mn\ \beta}_{\phantom{mn}\alpha}\Xsd_{mn}$
and
$\sigmabar^{mn\dot\alpha}_{\phantom{mn\alpha}\dot\beta}X_{mn}
=\sigmabar^{mn\dot\alpha}_{\phantom{mn\alpha}\dot\beta}\Xasd_{mn}$.
}
 then follows automatically from
the well-known self-duality property of the numerical tensor
 $\sigma_{mn}$.

The above construction does not  actually distinguish between the
gauge group $U(N)$ and $SU(N)$.
For, while the classical gauge field 
 constructed in this way is not automatically traceless, it
can be made so by a gauge transformation
$U \to U g^{\dagger}$, where $g^\dagger\in U(1)$. 
The distinction between $U(N)$ and $SU(N)$ only enters when
matter fields are coupled in. 
In the sections that follow, unless explicitly stated to the contrary, 	we
will work in the slightly more general $U(N)$ formalism; i.e, we
will not impose the 
tracelessness condition on adjoint matter fields.

In the next subsection we will count the independent degrees of freedom
of the ADHM configuration and confirm that it has precisely the
number of collective coordinates needed to describe a $k$-instanton
solution.

\subsec{Constraints, canonical forms, and collective coordinate counting}

We have seen that the ADHM construction for $SU(N)$ makes essential use
of matrices of various sizes:
$(N+2k)\times N$ matrices 
$U$,  $(N+2k) \times 2k$ matrices 
$\Delta$, $a$ and $b$, $k\times k$ matrices $f$, and
$2\times2$ matrices
 $\sigma^m_{ \alpha \dalpha}$, 
 $\sigmabar^{m \, \dalpha \alpha},$ $x_{\alpha\dalpha},$ etc. 
(Notice that when $N=2,$ the dimensionalities of $U$ and $\Delta$ differ
from the ``$SU(2)$ as $Sp(1)$'' formalism reviewed in Ref.~\dkmone.)
Accordingly, we introduce a variety
of index assignments:
$$\eqalignno{\hbox{Instanton number indices\ }[k]:\qquad&1\le i,j,l\cdots\le
k&\cr
\hbox{Color indices\ }[N]:\qquad&1\le u,v\cdots\le N&\cr
\hbox{ADHM indices\ }[N+2k]:\qquad&1\le \lambda,\mu\cdots\le N+2k&\cr
\hbox{Quaternionic (Weyl) indices\ }[2]:\qquad&\alpha,\beta,\dalpha,
\dbeta\cdots=1,2&\cr
\hbox{Lorentz indices\ }[4]:\qquad&m,n\cdots=0,1,2,3\hbox{ or }1,2,3,4&\cr
}$$
No extra notation is required for the $2k$ dimensional column index attached
to $\Delta,$ $a$ and $b$, since it can be factored as $[2k]=[k]\times[2]
=j\dbeta,$ etc., as in Eq.~\dlt. With these index conventions, Eq.~\dlt\ reads
\eqn\del{\Delta_{\lambda \, i \dalpha}(x)
\ =\ a_{\lambda \, i \dalpha}\ +\
b_{\lambda \, i}^{\beta}\,x_{\beta\dalpha}\ ,\qquad
\Deltabar^{\dalpha\lambda}_{ i  }(x)
\ =\ \abar^{\dalpha\lambda}_{ i }\ +\
\bar{x}^{\dalpha \alpha} \, \bbar^\lambda_{\alpha i}
}
while the factorization condition \dbd\ becomes
\eqn\fac{\Deltabar^{\dbeta\lambda}_{ j  }\,
\Delta^{}_{\lambda \, i \dalpha}\ =
 \  \delta^{\dbeta}_{\ \dalpha} \, {(f^{-1})}_{ij}\ .}
Combining Eqs.~\del-\fac, and noting that $f_{ij}(x)$ is arbitrary,
one extracts the three $x$-independent conditions on $a$ and $b$:
\eqna\cone
$$\eqalignno{
\abar^{\dalpha\lambda}_{ i }\, a_{\lambda  j \dbeta}\
&= \ (\hf \abar a)^{}_{ij} \ \delta^{\dalpha}_{\ \dbeta} 
\ \propto \ \delta^{\dalpha}_{\ \dbeta} &\cone a\cr
\abar^{\dalpha\lambda}_{ i  }\, b_{\lambda  j}^{\beta}\
&= \ 
\bbar^{\beta\lambda}_{ i} \, a_{\lambda  j}^\dalpha
&\cone b\cr
\bbar^\lambda_{\alpha i} \, b_{\lambda  j}^{\beta}\
&= \ (\hf \bbar b)^{}_{ij} \ \delta_{\alpha}^{\ \beta} 
\ \propto \ \delta_{\alpha}^{\ \beta}\ . &\cone c}$$
These three conditions are known as the ADHM constraints
\refs{\CGTone,\CWS}.

The elements of the matrices $a$ and $b$ comprise the collective coordinates
for the \hbox{$k$-instanton} gauge configuration. Clearly the number
of independent such elements grows as $k^2,$ even after accounting
for the constraints \cone{}. In contrast, the number of physical
collective coordinates should equal $4Nk$ which scales
 linearly with $k$.{$\,$}\foot{To see this, consider the limit of $k$
far-separated (distinguishable) instantons; each individual
instanton is then described
by four positions, one scale size, and $4N-5$ iso-orientations, totaling
$4N$ collective coordinates. Usually, as in \refs{\BCGW,\CWS}, 
the number of collective coordinates of the \hbox{$k$-instanton}
is quoted  as $4Nk-N^2+1$ for $k\geq N/2$ and
$4k^2+1$ for $k< N/2$. These formulae represent only true collective
coordinates, that is,  excluding the global gauge rotations
of the $k$-instanton configuration. In contrast,
in our counting we include such global rotations since they appear
 in our $k$-instanton measure. Then the total number of 
collective coordinates is $4Nk$.}
 It follows that
$a$ and $b$ constitute a highly redundant set. Much of this redundancy
can be eliminated by noting that the ADHM construction is
unaffected by $x$-independent transformations of the form
\eqn\tra{\eqalign{\Delta_{\sst [N+2k] \times [k] \times [2]}
 \ &\to \ \Lambda_{\sst [N+2k] \times \underline{[N+2k]} } 
 \ 
\Delta_{\sst \underline{[N+2k]} \times \underline{\sst [k]} 
\times [2] }
 \ B^{-1}_{\sst \underline{[k]} \times {[k]} }
\cr
U_{\sst [N+2k] \times [N]} \ &\to \
\Lambda_{\sst [N+2k] \times \underline{[N+2k]} } \ 
U_{\sst \underline{[N+2k]} \times [N]}
\cr
f_{\sst [k] \times [k]} \ &\to \
B_{\sst [k] \times \underline{[k]} } \ 
f_{\sst \underline{[k]} \times \underline{[k]}} \
B^\dagger_{\sst \underline{[k]} \times [k] }}}
provided $\Lambda \in U(N+2k)$ and $B \in Gl(k, {\bf C})$.
(These are in addition to the usual space-time gauge symmetries
reviewed in Sec.~6 of \dkmone.)
Exploiting these symmetries, one can choose
a representation in which $b$ assumes a simple  canonical
form \CGTone: 
\eqn\canform{b_{\sst [N+2k]\times [2k]} \ = \ 
\pmatrix{ 0_{\sst [N]\times [2k]} \cr \cr 1_{\sst [2k]\times [2k]}}   
\quad, \qquad
a_{\sst [N+2k]\times [2k]} \ = \ 
\pmatrix{ w_{\sst [N]\times [2k]} \cr \cr a'_{\sst [2k]\times [2k]}}
}

We can make this canonical form a little more explicit with a 
convenient representation of the index set  $[N+2k]$.
We decompose each ADHM index $\lambda\in[N+2k]$  into\foot{The Weyl
index $\beta$ in this decomposition is raised and lowered with the
$\epsilon$ tensor as always \WB, whereas for the $[N]$ and $[k]$ indices
$u$ and $l$ there is no distinction between upper and lower indices.}
\eqn\rplam{\lambda = u + l\beta\ ,\quad1\le u\le N\ ,\quad1\le l\le k\ ,
\quad\beta=1,2\ .  }
In other words, the top $N\times2k$ submatrices in Eq.~\canform\ have rows
indexed by $u\in[N],$ whereas the bottom $2k\times2k$ submatrices have
rows indexed by the pair $l\beta\in[k]\times[2].$ Equation \canform\
then becomes
%
%
\eqna\aaa
$$\eqalignno{a_{\lambda \, i \dalpha}\,&=
 \, a_{(u+l\beta) \, i \dalpha}\, = \, 
w_{u  i \dalpha}\,+ \, (a'_{\beta \dalpha})^{ }_{li} \,=\,
\pmatrix{  w_{u \, i \dalpha}\cr \cr 
(a'_{\beta \dalpha})^{ }_{li}}_{\phantom{q}}
\ ,&\aaa a\cr
\abar^{\dalpha\lambda}_{i} \, &= \, \abar_{i}^{\dalpha\,
(u+l\beta)}
 \, = \,
\wbar^\dalpha_{iu}+(\abar^{\prime\dalpha\beta})^{ }_{il}\,=\,
\big(\wbar^\dalpha_{i u}\ ,\ (\abar^{\prime\dalpha\beta})^{}_{il} 
\big)^{\phantom{T} }_{\phantom{q}}\ ,
&\aaa b\cr
b_{\lambda \, i}^\alpha\,&=\,b_{(u+l\beta)\, i}^\alpha\, = \,
\delta_\beta^{\ \alpha}\delta^{}_{li}\,=\,
\pmatrix{  0\cr \cr\delta_{\beta}^{\ \alpha}\, 
\delta_{li}^{} }^{\phantom{T}}_{\phantom{q}}
\ , &\aaa c\cr
\bbar_{\alpha i}^\lambda \, &= \, \bbar_{\alpha i}^{u+l\beta} \, = \,
\delta_\alpha^{\ \beta}\,\delta^{}_{il}\,=
\big(0 \ ,\ \delta^{\ \beta}_{\alpha}\, \delta_{il}^{} 
\big)^{\phantom{T}}\ .&\aaa d\cr
}$$
With $a$ and $b$ in the canonical form \aaa{}, 
the third ADHM constraint \cone c is
 satisfied automatically, while the remaining constraints \cone{a,b}
boil down to:
\eqna\fcone
$$\eqalignno{\trtwo ( \tau^c \abar a)_{ij} \ &= \ 0
&\fcone a\cr
(a^{\prime m})_{ij}^\dagger  \ &= \ a^{\prime m}_{ ij} \ .
&\fcone b\cr}$$
In Eq.~\fcone a we have contracted $\abar a$ with the three Pauli matrices
$({\tau^c})_{\ \dbeta}^{\dalpha}$, while in Eq.~\fcone b we have decomposed
 $(a'_{\beta \dalpha})_{li}$ and $(\abar^{\prime\dalpha\beta})_{il}$
in our usual quaternionic basis of spin matrices: 
\eqn\dec{(a'_{\beta \dalpha})^{}_{li} \ = \ 
(a'_m)^{}_{li} \ \sigma^m_{ \beta \dalpha} \ , \quad
(\abar^{\prime\dalpha \beta})^{}_{il} \ = \
(a'_m)^{}_{il} \ \sigmabar^{m \, \dalpha \beta}\ .}

Note that the canonical form for $b$ given in Eq.~\aaa{} 
is  preserved by a $U(k)$ subgroup
of the \hbox{$U(N+2k)\times Gl(k,{\bf C})$} symmetry group
\tra, namely:
\eqn\res{
\Delta_{\sst [N+2k]\times [2k]} \ \to \ 
\pmatrix{ 1_{\sst [N]\times [N]} & 0_{\sst [2k]\times [N]} \cr \cr 
0_{\sst [N]\times [2k]}  & {\cal R}^\dagger_{\sst [2k]\times [2k]} } 
\ \Delta_{\sst [N+2k]\times [2k]} \ {\cal R}_{\sst [2k]\times [2k]}
}
where ${\cal R}_{\sst [2k]\times [2k]} 
 =  R_{ij} \ \delta_{\ \dalpha}^\dbeta$
and $R_{ij} \in U(k)$. In terms of $w$ and $a'$, this residual
transformation acts as
\eqn\restw{w_{u  i}^{\dalpha}\, \ \to \ w_{u  j}^{\dalpha}\, R_{ji}
\ , \quad 
(a'_{\beta \dalpha})_{ij} \ \to \ R^\dagger_{il} \ 
(a'_{\beta \dalpha})_{lp}\ R_{pj}
}
It follows that
 the physical moduli space, ${\rm M}^k_{\rm phys}$, 
of inequivalent self-dual gauge configurations in the topological
sector $k$ is the
quotient of the space ${\rm M}^k$ of all 
solutions of the ADHM canonical constraints \fcone{},
by this residual symmetry group $U(k)$: 
\eqn\modspace{{\rm M}^k_{\rm phys} \  =\ {{\rm M}^k \over U(k)}\ .}

Finally we can count the 
independent collective coordinate degrees of freedom of the ADHM 
multi-instanton. A general complex matrix 
$a_{\sst [N+2k]\times [2k]}$ has $4k(N+2k)$ real degrees of freedom.
The two ADHM conditions \fcone{a,b} impose
$3k^2$ and $4k^2$ real constraints, respectively, while modding out
by the residual $U(k)$ symmetry removes another
 $k^2$ degrees of freedom. In total we therefore have
\eqn\dofb{
4k(N+2k) \ - \ 3k^2 \ - \ 4k^2 \ - k^2 \ = \ 4Nk }
real degrees of freedom, precisely as required. Of these, the four real
degrees of freedom $X_{\alpha\dalpha}=X_m\sigma^m_{\alpha\dalpha}$
corresponding to
\eqn\transmo{a_{\lambda i\dalpha}\ =\ b^\alpha_{\lambda i}\,X_{\alpha\dalpha}}
are the translational collective coordinates, as is obvious from Eq.~\del.

\subsec{The multi-instanton in singular gauge}

Although this is rarely needed in practice,
 we can determine the  instanton $v_m$ more 
explicitly.
Let us solve for $U$, and hence $v_m$ itself, in terms of $\Delta$.
It is convenient to make the decomposition:
\eqn\dcmp{U_{\sst [N+2k]\times [N]} \ = \
\pmatrix{ V_{\sst [N]\times [N]} \cr \cr U'_{\sst [2k]\times [N]}}
 \ , \quad
\Delta_{\sst [N+2k]\times [2k]} \ = \ 
\pmatrix{ w_{\sst [N]\times [2k]} \cr \cr 
\Delta'_{\sst [2k]\times [2k]}}
}
Then from the completeness condition \cmpl\ one finds
\eqn\cmpag{
 V_{\sst [N]\times [N]} \ \Vbar_{\sst [N]\times [N]} \ = \ 
1_{\sst [N]\times [N]} \ - \ 
w_{\sst [N] \times \underline{[k]} \times \underline{[2]}} \ 
f_{\sst \underline{[k]} \times \underline{[k]}} \ 
\wbar_{\sst \underline{[2]} \times \underline{[k]} \times [N]}
}
For any $V$ that solves this equation, one can find another by
right-multiplying it by a $U(N)$ matrix. A specific choice of 
$V$ corresponds to fixing the gauge. The ``instanton singular
gauges'' correspond to taking any one of the $2^N$ choices of
matrix square roots:
\eqn\sga{V \ = \ (1 \ - \ w f \wbar)^{1/2}  }
Next, $U'$ in Eq. \dcmp\ is determined in terms of
$V$ via
\eqn\prim{
U' \ = \ -\Delta' f \wbar \Vbar^{-1}  }
which likewise follows from Eq.~\cmpl.

Equations \sga\ and \prim\ determine $U$ in \dcmp,
and hence the gauge field $v_m$ via Eq. \vdef.
We list for later use the leading large-$|x|$ asymptotic behavior of
several key ADHM quantities, assuming instanton singular gauge
\sga:
\eqna\asymadhm
$$\eqalignno{\Delta\ &\rightarrow\ bx\ ,&\asymadhm a
\cr f_{kl}\ &\rightarrow\ {1\over|x|^2}\,\delta_{kl}\ ,&\asymadhm b
\cr U'\ &\rightarrow\ -{1\over|x|^2}\,x\,\wbar\  ,&\asymadhm c
\cr V\ &\rightarrow\ 1_{\sst [N] \times [N]} \ .&\asymadhm d}$$

As an example, let us verify that
 the usual $1$-instanton solution
\BPST\  follows from this general formalism. We adopt
the canonical form \aaa{} and set the 
instanton number $k=1$, thus dropping the $i,j$ indices. Contrary to the
``$SU(2)$ as $Sp(1)$'' treatment of Ref.~\dkmone, now the ADHM constraints
\fcone{} do not disappear in the 1-instanton sector. Instead,
Eq.~\fcone b says that $a'_m$ is real,
\eqn\xcon{a'_m \ \equiv \ X_m \ \in \ \bigR\ , }
after which Eq.~\fcone a collapses to
\eqn\wcon{\wbar^\dalpha_u \, w_{u \dbeta} \ = \ 
\rho^2\,\delta^\dalpha_{\ \dbeta}  \ .}
\eqna\omom
(The quantities $\rho$ and $-X^n$ will be identified with the instanton
scale size and space-time position, respectively.)
It follows that the two complex $N$-vectors $\omega_u^{(\dalpha)}$
defined by $\omega_u^{(\dalpha)}=w_{u\dalpha}/\rho$ are orthonormal.
It is convenient to put them in the form:\foot{As a quick
check, note that $w_{u\dalpha}$ has $4N$ real degrees of freedom, of which
three are eliminated by Eq.~\wcon. This agrees with the counting
from Eq.~\omom{}: the coset element $\Omega$ has
 $N^2-(N-2)^2= 4N-4$ real degrees of freedom, and the scale size
$\rho$ has one, for a total of $4N-3$ in both cases. Adding in
the four translational degrees of freedom $X^m$ from Eq.~\xcon, and
subtracting the residual $U(1)$ from Eq.~\restw, makes a grand total of $4N$
independent collective coordinates, in accord with the counting \dofb.}
$$\eqalignno{&\omega_{u}^{(\dalpha)}
\ = \ \Omega_{\sst [N] \times [N]} \ 
\pmatrix{0_{\sst [N-2] \times [2]} \cr \cr
  1_{\sst [2]\times [2]}} \ , \qquad 
\Omega \in {U(N) \over U(N-2)}\ .&\omom{}}$$
Setting $\Omega=1$ initially, we find for $\Delta$ and $f\,$:
\eqn\dlsi{\Delta_{\sst [N+2]\times [2]} \ = \ 
\pmatrix{ 0_{\sst [N-2] \times [2]} \cr 
  \rho \cdot 1_{\sst [2]\times [2]} \cr
  y_{\sst [2]\times [2]} }\ ,\qquad
f  \ = \ {1 \over y^2 + \rho^2} \ , }
with $y_{\alpha\dalpha} =  (x+X)_{\alpha\dalpha}.$ 
Equations \sga-\prim\ then amount to
\eqn\veq{
V_{\sst [N]\times [N]}  \ = \ 
\pmatrix{ 1_{\sst [N-2]\times [N-2]} & 0 \cr
0 & \left( {y^2 \over y^2 + \rho^2} \right)^{1/2} 
1_{\sst [2]\times [2]}  } }
and
\eqn\ueq{U'_{\sst [2]\times [N]} \ = \ 
\pmatrix{ 0_{\sst [2]\times [N-2]} &\ , \ &
-\left( {\rho^2 \over y^2 (y^2+ \rho^2)} \right)^{1/2} 
y_{\sst [2]\times [2]}  }\ .}
The gauge field follows from Eq. \vdef\ as in Eq.~(6.3) of Ref.~\dkmone:
\eqn\sutin{v_m \ = \ \pmatrix{ 0 & 0 \cr 0 & v_m^{\sst SU(2)} } }
where
$v_m^{\sst SU(2)}$ is the standard singular-gauge $SU(2)$ instanton \tHooft\
with  space-time position $-X_n\,$, scale-size $\rho$, and
in a fixed ``reference'' iso-orientation:
\eqn\singin{
 v_m^{\sst SU(2)} (x)\ = \ {\rho^2 \ \etabar^a_{mn}\ (x^n+X^n) \ \tau^a   
\over (x+X)^2 \ ((x+X)^2 + \rho^2)}
\ . }
For a general $\Omega$ we obtain instead
%
%
\eqn\inss{v_m \ = \ \Omega \
\pmatrix{ 0 & 0 \cr 0 & v_m^{\sst SU(2)} } \bar{\Omega} 
\ , \qquad 
\Omega \in {U(N) \over U (1) \times U(N-2)}\ .
}
The extra $U(1)$ in the denominator in Eq.~\inss\ is  the
residual symmetry \restw.

%
%
%

%
%

\newsec{Construction of the ADHM Supermultiplet}


\subsec{The case of $\N=1$ \susy}

In an $\N=1$ supersymmetric theory the gauge field $v_m$ is accompanied 
by a gaugino $\lambda$. In the ADHM background there are 
non-trivial solutions to the covariant Weyl equation $\Dbarslash \lambda =0$.
By the index theorem, the zero modes of $\Dbarslash$ should comprise 
$2Nk$ independent Grassmann degrees of freedom. 
As discussed in \NSVZ\ in the 1-instanton context, these zero modes
can be considered the $\N=1$ superpartners of the instanton.
Explicit expressions for the adjoint fermion zero modes in the ADHM
background were first obtained in \CGTone. In our notation they
read (cf.~Eq. (7.1) of \dkmone):
\eqn\lam{(\lambda_\alpha)_{uv} \ = \ 
\Ubar^\lambda_{u} \M_{\lambda i} f_{ij}\bbar^\rho_{\alpha j}U_{\rho v} \ - \ 
\Ubar^\lambda_{u} 
b_{\lambda i\alpha}\, f_{ij} \Mbar^\rho_{j} U_{\rho v} \ . }
Here $\M_{\lambda i}$ and $\Mbar^\rho_{j}$ are constant $(N+2k)\times k$ 
and $k\times (N+2k)$ matrices of Grassmann collective coordinates;
they can be viewed as either  
two  real Grassmann matrices or as two complex 
Grassmann matrices which are Hermitian conjugates of one another.

{}From Eq.~\lam\ we calculate (as in Sec.~7.2 of \dkmone):
\eqn\zmid{\Dbarslash^{\dalpha\alpha}
\lambda_\alpha \ =\ 
2\Ubar b^\alpha\,f\big(\Deltabar^{\dalpha}\M+
\Mbar\Delta^\dalpha\big)f\,\bbar_\alpha\, U}
Hence the condition for a gaugino zero mode 
is the following two sets of
linear constraints on $\M$ and $\Mbar$ which ensure that the right-hand side
vanishes (expanding $\Delta(x)$ as $a+bx$) \CGTone:
\eqna\zmcon
$$\eqalignno{\Mbar^\lambda_{i} \, a_{\lambda j\dalpha}
 \ &= \ - \abar^\lambda_{i\dalpha} \, \M_{\lambda j}
\ ,&\zmcon a\cr
\Mbar^\lambda_{i} \, b_{\lambda j}^\alpha \ &= \  \bbar_{i}^{\alpha\lambda}
 \, \M_{\lambda j}\ .
&\zmcon b\cr}$$
In a formal sense discussed in Sec.~6 below, these fermionic constraints
are the ``spin-$1/2$'' superpartners of the original ``spin-1''
 ADHM constraints \cone{a,b}, respectively.
Note further that Eq.~\zmcon b
 is easily solved when $b$ is in the canonical form
\aaa{}. 
With the ADHM index decomposition \rplam, we set
\eqn\mrep{\M_{\lambda i} \ \equiv \ \M_{(u+l\beta)\, i} \ = \
\pmatrix{ \mu_{u i} \cr \cr (\M'_\beta)_{li} }   \ ,\quad
\Mbar^\lambda_{i} \ \equiv \ {\Mbar_i}^{u+l\beta} \ = \ 
\left( \mubar_{i u} \ ,\ (\Mbar^{\prime\beta})_{il} \right)\ .}
Equation \zmcon b then collapses to
\eqn\mctw{ \Mbar^{\prime\alpha}\ = 
\M^{\prime\alpha}}
which allows us to eliminate $\Mbar'$ in favor of $\M'$.

Counting the number of degrees of freedom, one finds $2k(N+2k)$ 
real Grassmann parameters in  $\M$ and $\Mbar$, subject to $2k^2$ 
constraints from 
each of Eqs.~\zmcon{a,b}, for a net of $2Nk$ gaugino
zero modes as required. Of these, two Weyl spinor zero modes
can be distinguished, namely
\eqn\susymo{\M_{\lambda i}\ =\ 4b_{\lambda i}^\beta\,\xi_\beta\ ,\qquad
\Mbar^\lambda_i\ =\ 4\bbar^\lambda_{i\beta}\,\xi^\beta}
and
\eqn\suconmo{\M_{\lambda i}\ =\ i\,a_{\lambda i\dalpha}\,\etabar^\dalpha\ ,
\qquad \Mbar^\lambda_i\ =\ -i\,\abar^{\dalpha\lambda}_i\,
\etabar_\dalpha\ ,}
where $\xi_\beta$ and $\etabar^\dalpha$ are arbitrary spinor parameters.
These are the so-called ``supersymmetric'' and ``superconformal''
zero modes, respectively \NSVZ; they satisfy the
 fermionic constraints \zmcon{} by virtue of the ADHM constraints
\cone{a,b}. 

As for the remaining $2Nk-4$ modes, the simplest case
to study is
the 1-instanton sector, $k=1$, with the instanton oriented as in 
Eq.~\omom{}, and with $\Omega$ set to unity. Apart from the \susic\
and superconformal modes \susymo-\suconmo, there are $2N-4$ additional
fermionic zero modes which are the   superpartners to gauge orientations
\Cordes. They are constructed by setting
 $\M'=0$ and also $\mu_u=0$ for $u=N-1$ or $N$, with
arbitrary choices for $\mu_u$ for $u\le N-2$; by inspection,
these satisfy the constraints \zmcon{}. Turning on the orientation
matrix $\Omega$ as in 
Eq.~\omom{} simply rotates these choices of $\M$ by $\Omega$.
For $k=1$ all these modes correspond to Lagrangian symmetries, but
in the higher-instanton sectors there are $2N(k-1)$ ``relative''
fermionic zero modes  which do not have such an interpretation.

\subsec{The  case of ${\cal N}=2$ \susy}

Next we turn to the $\N=2$ case. 
The particle content of $\N=2$ \susic\ Yang-Mills theory
comprises, in addition to the gauge field $v_m$ and gaugino $\lambda_\alpha$
discussed above, a Higgsino $\psi_\alpha$ and a
 complex Higgs boson $A$. All transform in the adjoint representation
of the $U(N)$ gauge group.

\it Adjoint Higgsino zero modes. \rm

The fermion zero modes of the Higgsino
$\psi$ are defined in identical fashion to
 the gaugino, Eqs.~\lam-\mctw:
\eqn\psizm{(\psi_\alpha)_{uv} \ = \ 
\Ubar^\lambda_{u} \N_{\lambda i} f_{ij}\bbar^\rho_{\alpha j}U_{\rho v} \ - \ 
\Ubar^\lambda_{u} 
b_{\lambda i\alpha}\, f_{ij} \Nbar^\rho_{j} U_{\rho v} \ . }
%
The Grassmann-valued collective coordinate matrices 
 $\N_{\lambda i}$ and $\Nbar_{j}^\rho$ are subject to the same linear
 constraints as $\M$ and $\Mbar\,$,
\eqna\pscon
$$\eqalignno{\Nbar^\lambda_{i} \, a_{\lambda j\dalpha}
 \ &= \ - \abar^\lambda_{i\dalpha} \, \N_{\lambda j}
\ ,&\pscon a\cr
\Nbar^\lambda_{i} \, b_{\lambda j}^\alpha \ &= \  \bbar_{i}^{\alpha\lambda}
 \, \N_{\lambda j}\ ,
&\pscon b\cr}$$
and are likewise decomposed as
\eqn\nrep{\N_{\lambda i} \ \equiv \ \N_{(u+l\beta)\, i} \ = \
\pmatrix{ \nu_{u i} \cr \cr (\N'_\beta)_{li} }   \ ,\quad
\Nbar^\lambda_{i} \ \equiv \ {\Nbar_i}^{u+l\beta} \ = \ 
\left( \nubar_{i u} \ ,\ (\Nbar^{\prime\beta})_{il} \right)\ ,}
with
\eqn\nnctw{ \Nbar^{\prime\alpha}\ = 
\N^{\prime\alpha}}
in the canonical basis for $b$.

\it The adjoint Higgs boson. \rm

In the ADHM background the
complex scalar field $A$ satisfies the classical Euler-Lagrange
equation\foot{As in \refs{\dkmone,\dkmfour}, in our conventions 
the only anti-Hermitian field
is the gauge field $v_m$, while all other component fields are
 Hermitian.}
\eqn\Higgseq{\D^2 A\ =\ \sqrtwo i\,[\,\lambda\,,\psi\,]}
where $\D^2$ is the covariant Klein-Gordon operator in the
multi-instanton  background, and $\lambda$ and $\psi$ are given
by \lam\ and \psizm, respectively.
The boundary condition
\eqn\vevs{
A(x) \to {\rm diag}(\vhiggs_{\sst 1}, \ldots, \vhiggs_{\sst N}) \ , \qquad
|x|\ \to\  \infty}
specifies the complex VEVs $\vhiggs_u$. These
are not constrained by $\N=2$
supersymmetry, and should be viewed as $N$ free complex parameters (``moduli'')
characterizing a given $\N=2$ theory.

The construction of the classical Higgs $A$ is the generalization
of the expression obtained in Secs. 7.2-7.3 of \dkmone, and
goes as follows. $A$ has the additive form 
\eqn\Aonedef{i\,A \ =\
{1\over2\sqrtwo}\,\Ubar\,\big(\N f\Mbar
-\M f\Nbar\big)U\ +\  \Ubar \,\A \,U \ .}
Here $\A$ is a block-diagonal constant $(N+2k)\times (N+2k)$  matrix,
\eqn\blockdiag{\A_\lambda^{\ \mu}\ \equiv \ 
\A_{u+l\alpha}^{\  v+m\beta}
\ =\
\pmatrix{\homoa_{uv} &0 \cr {}\cr
0 &(\Atot)_{lm}\,\delta_\alpha^{\ \beta}}\ ,}
where the $N\times N$ matrix $\homoa$ is just $i$ times the VEV matrix,
\eqn\vevbcagain{{\homoa}_{uv}
\ =\ i \ {\rm diag}(\vhiggs_{\sst 1}, \ldots, \vhiggs_{\sst N}) \ .}
The $k\times k$ anti-Hermitian matrix $\Atot$ is 
defined as the
solutions to the inhomogeneous linear  equation
\foot{In the remainder of the paper we distinguish two different kinds of
Hermitian conjugation. The first type, denoted by a dagger, does not
turn fields into anti-fields, nor does it complex conjugate the VEVs.
Thus:
${\homoa}^{\dagger}_{uv}
 = -i\,  {\rm diag}(\vhiggs_{\sst 1}, \ldots, \vhiggs_{\sst N}).$
The second (standard) type of Hermitian conjugation, denoted by an
overbar,
does interchange fields and anti-fields and also complex-conjugates the VEVs.
Thus:
${\barhomoa}_{uv}
 = -i  \,{\rm diag}(\vhiggsbar_{\sst 1}, \ldots, \vhiggsbar_{\sst N}).$
For the remainder of this section, Hermitian conjugation is always of
the first type.}
\def\Atotdagger{{\cal A}_{\rm tot}^\dagger}
\eqn\thirtysomething{\bigL\cdot\Atot\ =\ \Lambda+\Lambda_f\ ,
\quad \Atotdagger \ = \ -\Atot}
where $\Lambda$ and $\Lambda_f$ are the $k\times k$ anti-Hermitian matrices
\def\Lambdabar{\bar\Lambda}
\eqn\Lambdabardef{\Lambda_{ij}\ =\ \wbar^\dalpha_{i u} \ \homoa_{uv} \ 
w_{v j\dalpha} \ , 
\quad  \Lambda^\dagger \ = \ -\Lambda }
and
\eqn\newmatdef{ (\Lambda_f)_{ij}\ =\ {1\over2\sqrtwo}\,
\big(\,\Mbar\N -\Nbar\M \,\big)_{ij}\ , \quad
\Lambda_f^\dagger \ = \ -\Lambda_f }
$\bigL$ is a linear operator that maps the space of $k\times k$ 
 scalar-valued anti-Hermitian matrices onto itself. Explicitly,
if $\Omega$ is such a matrix, then $\bigL$ is defined as \dkmone
\eqn\bigLreally{\bigL\cdot\Omega\ =\ 
\hf\{\,\Omega\,,\,W\,\}\,-\,\hf\trtwo\big(
[\,\abar'\,,\,\Omega\,]a'-\abar'[\,a'\,,\,\Omega\,]\big)}
where 
$W$ is the Hermitian  $k\times k$ matrix
\eqn\Wdef{ W_{ij}\ =\ \wbar^\dalpha_{i u} \  
w_{u j\dalpha} \ , 
\quad  W^\dagger \ = \ W}
{}From Eqs.~\thirtysomething-\Wdef\ one sees that $\Atot$
transforms in the adjoint representation of the residual $U(k)$ \res\
(i.e., like
$a',$ $\M'$ and $\N'$). 

Defined in this way, the Higgs field $A$ correctly
satisfies the equation \Higgseq; see Sec.~7 of Ref.~\dkmone\ for calculational
details.
We also note that the constraints \cone{a,b}, \zmcon{a,b}, \pscon{a,b},
and \thirtysomething\ may be thought of, respectively, as the
``spin-1,'' ``spin-$1/2$,'' ``spin-$1/2$,'' and
``spin-0'' components of an $\N=2$ supermultiplet of constraints
\dkmfour. We will exploit this observation in Sec.~6 below, when we construct
the collective coordinate integration measure.

\newsec{Realization of the Supersymmetry Algebra}

\subsec{The case of $\N=1$ \susy}

Here we discuss the \susic\ transformation
properties of the collective coordinate
matrices $a$ and $\M$,
following the formalism developed in \dkmfour. The philosophy is as
follows \NSVZ.
 As the relevant
field configurations $v_m$ and $\lambda_\alpha$
obey equations of motion which are manifestly
supersymmetric, any non-vanishing action of 
the  supersymmetry generators on a 
particular classical solution necessarily yields another solution. It follows
that the ``active''
supersymmetry transformations of the fields must be equivalent (up  
to a
gauge transformation) to certain ``passive'' transformations of 
the $4Nk$ independent bosonic and $2Nk$ independent
fermionic collective coordinates which parametrize
the superinstanton solution. As originally noted in \NSVZ\ in the 1-instanton
context, 
physically relevant quantities such as the
the superinstanton action must be constructed out of
\susic\ invariant combinations of the collective coordinates.   

As explained in \dkmfour, the 
supersymmetry algebra can actually be realized  
directly as transformations of  collective
coordinates $a$ and $\M$
{\it before} implementing the respective algebraic constraints \cone{}
and \zmcon{}.
The analysis is identical to Sec.~2 of \dkmfour\ and need not be
repeated here. One finds
 that under
an infinitesimal \susy\ transformation 
$-i\xi Q+i\xibar\Qbar,$
the collective coordinates
 transform as\foot{Here and in the $\N=2$ case to follow,
we redefine the infinitesimal \susy\ parameters of 
Refs.~\refs{\dkmone,\dkmfour} as
 $\xi\rightarrow-i\xi,$ $\xibar\rightarrow i\xibar.$}
\eqna\susyalgebra
$$\eqalignno{\delta a_{\dalpha}\ &=\ i\xibar_{\dalpha}\M\ ,\qquad\ \ \,
\delta \abar^{\dalpha}\ =\ -i \Mbar \xibar^{\dalpha}
&\susyalgebra a \cr
\delta\M\ &=\ -4b^{\alpha}\xi_{\alpha}\ ,\qquad
\delta\Mbar\ =\ -4\xi^{\alpha}\bbar_\alpha
&\susyalgebra b \cr}$$

\subsec{The case of $\N=2$ \susy}

\eqna\susyalgebratwo
As in the $\N=1$ case, the $\N=2$ \susy\ algebra may likewise 
be realized directly
on the unconstrained multi-instanton collective coordinates. As 
in Sec.~2 of \dkmfour,
under the action of $\sum_{i=1,2}
\,(-i\xi_iQ_i+i\xibar_i\Qbar_i)\,$ one has:
$$\eqalignno{
\delta a_{\dalpha}\ &=\ i\xibar_{1\dalpha}\M
+i\xibar_{2\dalpha}\N \ ,\qquad
\qquad\ \,
\delta \abar^{\dalpha}\ =\ -i \Mbar \xibar_{1}^{\dalpha}
-i\Nbar \xibar_{2}^{\dalpha}
&\susyalgebratwo a \cr
\delta\M\ &=\ -4b^{\alpha}\xi_{1\alpha}
-i2\sqrtwo\C_{\dalpha}
\xibar_2^\dalpha  \ ,\qquad
\delta\Mbar\ =\ -4\xi_{1}^{\alpha}\bbar_\alpha
+i2\sqrtwo \xibar_{2\dalpha} \C^{\dagger \dalpha}
&\susyalgebratwo b \cr
\delta\N &=\ -4b^{\alpha}\xi_{2\alpha}+i2\sqrtwo \C_{\dalpha}
\xibar_1^\dalpha \ ,\qquad
\,
\delta\Nbar\ =\ -4\xi_{2}^{\alpha}\bbar_\alpha
-i2\sqrtwo \xibar_{1\dalpha} \C^{\dagger \dalpha}
&\susyalgebratwo c \cr
}$$
Here $\C_{\dalpha}$ is the $(N+2k)\times k$ spinor-valued matrix
\eqna\Cdef
$$\eqalignno{\C_{\lambda \ i\dalpha}\ &\equiv \ \C_{(u+l\beta) \, i\dalpha}
\ = \ 
\pmatrix{\homoa_{uv} w_{v i\dalpha}-w_{uj\dalpha}(\Atot)_{ji}
\cr
\cr
\big[\,\Atot\,,\,a'_{\beta \dalpha}\,]_{li}^{}}
 \ ,&\Cdef a\cr
\C^{\dagger \dalpha} \ &= \ 
\big(\,\Atot \wbar^\dalpha - \wbar^\dalpha \homoa  \ , \ 
[\,\Atot\, , \abar^{\prime \dalpha}\,]\,\big) \ . &\Cdef b }$$
Direct calculation (see Appendix A of \dkmfour)
shows that $\Atot$, as defined in \thirtysomething\ above,
 is a supersymmetry invariant:
\eqn\atotr{\delta\Atot\ =\ 0\  }

\newsec{Construction of the Multi-Instanton Action}

\subsec{The case of $\N=1$ \susy}

In the absence of matter multiplets, the $k$-instanton action of $\N=1$
\susic\ $SU(N)$ Yang-Mills theory is simply $8\pi^2k/g^2,$ i.e., $k$
times the classical 1-instanton action. An interesting result can only
be obtained in the presence of a Higgs boson whose VEV explicitly breaks
the classical scale invariance of the theory. Let us start by considering
the simplest such theory, in which the gauge multiplet is minimally
coupled to a single fundamental chiral multiplet $Q=(q_u,\chi_u)$, where
$q_u$ is the Higgs, $\chi_u$ is the Weyl Higgsino, and the subscript
$u\in[N]$ indexes the fundamental representation. 

The fundamental fermion zero modes were originally constructed in \CGTone.
In our language, they read:
\eqn\fund{\chi^{\alpha}_{u }  \ = \
\bar{U}_{u\lambda}{b_{\lambda i}}^\alpha f_{ij}{\cal K}_{j}}
where $\alpha$ is a Weyl spinor index, and $\K_j$ is a Grassmann number
(as opposed to a Grassmann spinor). It is easily verified that $\chi$ so
defined
satisfies the covariant Weyl equation in the ADHM background,
\eqn\covWeyl{\Dbarslash\,\chi\ =\ 0\ .}
On the other hand, the fundamental Higgs $q_u$ satisfies an inhomogeneous
Euler-Lagrange equation,
\eqn\squarkeq{\D^2 q\ =\ -i\sqrtwo\lambda\chi}
together with the VEV boundary conditions 
\eqn\vevbcon{q_u\  \buildrel |x|\rightarrow\infty\over\longrightarrow
\ \qvac_u}
where $\qvac_u$ denotes the fundamental VEV. The right-hand side of
Eq.~\squarkeq\ is the product of the classical configurations
\lam\ and \fund, respectively. The general solution to Eqs.~\squarkeq-\vevbcon\
 is a straightforward exercise in ADHM algebra, and reads:
\eqn\squarkd{q_{u}\ =\ \bar{V}_{uv} \,\qvac_{v} \ +
\ {i\over2\sqrtwo}
\, \Ubar_{u \lambda}
\M_{\lambda i} f_{ij}\K_{j}  \ , }
generalizing Eq.~(5.10) of \dkmfour\ for the $SU(2)$ case. Here $V$,
defined in  Eq.~\dcmp\ above, is the upper $N\times N$ part of  the ADHM
 matrix $U$.

We can now construct the superinstanton action. The Maxwell term in
the component Lagrangian yields $8k\pi^2/g^2$ as always.
Following the method
of Refs.~\refs{\dkmone,\dkmfour},
the two other relevant terms of the component
Lagrangian, namely the Higgs kinetic energy and the Yukawa interaction, are
turned into a surface term with an integration by parts in the former
together with the Euler-Lagrange equation \squarkeq\
for the fundamental scalar.
 As per the divergence theorem, their contribution to the action
may then be extracted from the $1/x^3$ fall-off of $\D_{\sst\perp} q,$  
where
 the normal covariant derivative
$\D_{\sst\perp}$ is defined as $(x^m/\sqrt{|x|^2})\,\D_m\,$. With the 
help of the asymptotic formulae \asymadhm{}, one calculates
\eqn\funasym{\D_{\sst\perp} q_{u}\quad
 \buildrel |x|\rightarrow\infty\over\longrightarrow
\quad {1\over 2 |x|^3}\,
\big(
w_{\dalpha u i } \ \wbar^{\dalpha}_{iv} \ \qvac_{v}
\ -\ {i\over\sqrtwo}\,\mu_{ui}\ \K_{i}\big) \ , }
and hence
\eqn\funaction{\eqalign{S_{\sst \N=1 \ SQCD}^{k-\inst}\ =\ 
{8 k \pi^2 \over g^2} \ + \ 
\pi^2\, & \big(\,\qvac_{u}\,\qbvac_{v}\,
w^{}_{\dalpha v i } \, \wbar^{\dalpha}_{iu} 
\ -\ {i\over\sqrtwo}\,\qbvac_{u}\ \mu_{ui}\ \K_{i}\,\big)
}}
See Appendix C of \dkmfour, as well as Ref.~\oldyung, for the
analogous $SU(2)$ expressions.

This $k$-instanton formula, although written in ADHM collective coordinates,
 is nevertheless easily compared with the 1-instanton
expression for the action found in Ref.~\NSVZ: the first
term in parentheses is equivalent to $\sum_i\,|q|^2\,\rho_i^2,$
summed over the $k$ different instantons, where $q$ is the
fundamental VEV and $\rho_i$ is the
scale size of the $i$th instanton. Also the second term in parentheses
is the fermion bilinear necessary to promote this $\rho_i^2$ to
$(\rho^2_{\rm inv})_i$ where $\rho_{\rm inv}$ is the \susic\ invariant
scale size constructed in \NSVZ.
Independent of one's choice of collective coordinates, 
the presence of the VEV in the action \funaction\ gives a natural
cutoff to the integrations over instanton scale sizes \tHooft,
providing an infrared-safe application of instanton calculus.

The expressions given above may be immediately extended to phenomenologically
more interesting models with $N_F$ fundamental
flavors of Dirac fermions. In this
case the gauge multiplet is minimally coupled to $2N_F$ chiral superfields
$Q_f$ and $\Qtilde_f,$ $\,1\le f\le N_F,\,$ where $Q_f$ transforms in
the fundamental  and $\Qtilde_f$ in the conjugate-fundamental
representation of the gauge group. 
%
%
The classical moduli space of the theory in the Higgs phase 
is given by \refs{\IS,\Shifman}:
\eqn\qvev{
\qvac_{uf} \ = \ \pmatrix{
\vhiggs_1 & 0 & \ldots & 0 & \ldots & 0 \cr
0 & \vhiggs_2 & \ldots & 0 & \ldots & 0 \cr
\vdots & \vdots & \ddots  & \vdots  & \ddots  & \vdots \cr
0 & 0 & \ldots & \vhiggs_N & \ldots & 0 } \ , \quad
\qtvac_{fu} \ = \ \pmatrix{
\tilde{\vhiggs}_1 & 0 & \ldots & 0  \cr
0 & \tilde{\vhiggs}_2 & \ldots & 0  \cr
\vdots & \vdots & \ddots  & \vdots \cr
0 & 0 & \ldots &\tilde{\vhiggs}_N\cr
0 & 0 & \ldots & 0
}}
The VEV matrices   in Eq.~\qvev\ correspond to the cases
 $N_F \geq N$. The cases $N_F < N$ are similar except that the VEV
matrices have extra rows of zeroes rather than columns.
These VEVs are not all independent;
the D-flatness condition requires that for each value of $u$,
\eqna\dflg
$$\eqalignno{|\vhiggs_u|^2 \ &=\ 
 |\tilde{\vhiggs}_u|^2 + a^2 \ , \quad
N_F \geq N &\dflg a\cr
|\vhiggs_u|^2 \ &=\  |\tilde{\vhiggs}_u|^2 \ , \quad
\qquad\
N_F < N &\dflg b}$$
where $a^2$ is an arbitrary constant, independent of the color index $u$.

Now Eqs.~\fund\ and \squarkd\ generalize to
\eqn\fundredux{\chi^{\alpha}_{u f}  \ = \
\bar{U}_{u\lambda}{b_{\lambda i}}^\alpha f_{ij}{\cal K}_{jf}
\ ,\quad
\tilde{\chi}_{f\alpha u}  \ = \
\tilde{\cal K}_{fi} f_{ij}\bbar_{\alpha j \lambda} U_{\lambda u}
}
and
\eqn\squarkdredux{\eqalign{q_{uf}\ &=\ \bar{V}_{uv} \qvac_{vf} \ +
\ \textstyle{i\over2\sqrtwo}
\ \Ubar_{u \lambda}
\M_{\lambda i} f_{ij}\K_{jf}  \ , \cr
\tilde{q}_{fu}\ &=\ 
\qtvac_{fv} {V}_{vu}  \ -
\ \textstyle{i\over2\sqrtwo}
\ \Kt_{fi} f_{ij}
\Mbar_{j \lambda }U_{\lambda u} \ , }}
respectively, while the action becomes:
\eqn\funactionredux{\eqalign{S_{\sst \N=1 \ SQCD}^{k-\inst}\ =\ 
{8 k \pi^2 \over g^2} \ + \ 
\pi^2 & \big(\qvac_{uf}\qbvac_{fv}
w_{\dalpha v i } \ \wbar^{\dalpha}_{iu} 
\ -\ {i\over\sqrtwo}\,\qbvac_{fu}\ \mu_{ui}\ \K_{if}
\cr& +\ \qtbvac_{uf}\qtvac_{fv}
w_{\dalpha v i } \ \wbar^{\dalpha}_{iu} 
\ +\ {i\over\sqrtwo}\,\qtbvac_{uf}\ \Kt_{fi}\, \mubar_{iu}
\big)}}

As mentioned above, on general principle this action must be
a \susy\ invariant \refs{\NSVZ,\dkmfour}. The $\N=1$
\susy\ transformation
properties of the collective coordinate matrices $a$ and $\M$ (including
the submatrices $w$ and $\mu$) were constructed above, in Eq.~\susyalgebra{}.
To check the invariance of the expression \funactionredux, it is
necessary as well to derive the transformation properties for the
Grassmann collective coordinates $\K$ and $\Kt$ associated with the
fundamental fermions. As with the other collective coordinates, this
may be straightforwardly accomplished by equating ``active'' and ``passive''
\susy\ transformations on the Higgsinos $\chi$ and $\tilde\chi.$
In this way one obtains:
\eqn\deltaKnz{\delta\K_{if}\ =\ 
- 2 \sqrtwo \ \xibar_\dalpha \ \wbar^\dalpha_{iu}
\ \qvac_{uf}  \ ,\quad
\delta\Kt_{fi}\ =\ - 2 \sqrtwo \ \qtvac_{fu} \ w_{\dalpha u i}
\ \xibar^\dalpha  \ .}
It is now easily checked that the action \funactionredux\ is  invariant
under the \susy\ transformations \susyalgebra{} and  \deltaKnz.

In the next subsection, we turn our attention to  the multi-instanton
action on the Coulomb branch of $\N=2$ SQCD. We will see that the
$\N=1$ action \funactionredux\ possesses two simplifying
properties that the $\N=2$
action does not. 
First, Eq.~\funactionredux\
 has the form of a disconnected sum of $k$ single instantons;  
with the choice of these ADHM coordinates there is no interaction between them.
Second, the only gaugino modes that are lifted (i.e., that appear
in the action) are those associated
with the top elements $\mu$ and $\mubar$ of the collective coordinate  
matrices
$\M$ and $\Mbar$. This leaves $\CO(k)$ unlifted gaugino modes 
after one implements the fermionic
constraints \zmcon{}. This counting contrasts
sharply with the $\N=2$ theories in which the number of unlifted modes
is independent of the winding number $k$. 
Saturating each of these unlifted modes with an anti-Higgsino as per 
Affleck, Dine and Seiberg \ADS, one sees that
unlike the $\N=2$ theory, here the sectors of
different topological number cannot interfere with one another, since the
corresponding Green's functions
are distinguished by different (anti)fermion content.

\subsec{The case of $\N=2$ \susy}

Next we discuss the multi-instanton action in $\N=2$ SQCD. We start by
considering the case of pure Yang-Mills theory, $N_F=0.$ 
As above, the supersymmetric multi-instanton action
can be expressed as a surface term \dkmone:
\eqn\instaction{\eqalign{S_{\sst \N=2 \ SYM}^{k\hbox{-}\inst}\ 
&=\ \trN\int d^4x\,\Big(\ \hf\, v_{mn} v^{mn}\ 
-\ 2\,\D_m\Abar\D^m A\ +\
2\sqrtwo i\,[\,\Abar,\psi\,]\,\lambda\ \Big)
\cr&=\ {8k\pi^2\over g^2}\ 
-\ 
2\,\trN\int d^3{\rm S}\,\Abar\,\hat x_m\,\D^m A\ ,}}
where $\hat x_m=x_m/\sqrt{|x|^2}$ and S is the 3-sphere at 
infinity. The fields in Eq.~\instaction\ are assumed to be the
classical configurations constructed in Secs.~2-3 above;
the last equality follows from an integration by parts together with
the Euler-Lagrange equation \Higgseq. 
Evaluating the asymptotic value of $\Abar\,\hat x_m\,\D^m A$
with the help of \asymadhm{},
we obtain the expression for the $k$-instanton action:\foot{Note
 that  $\barhomoa$
and $\Lambdabar$
are 
Hermitian conjugations of the second type defined in footnote 8,
with complex conjugated VEVs.
 They are not to be confused with
$\homoa^\dagger$ and $\Lambda^\dagger$ in Sec.~3.2.}
\eqn\siact{\eqalign{S_{\sst \N=2 \ SYM}^{k\hbox{-}\inst} \ &= \ 
{8 k \pi^2 \over g^2} \ + \ 
8\pi^2 \ \wbar^\dalpha_{iu} \barhomoa_{uu} \homoa_{uu}
 w_{ui\dalpha} \ - \ 
8\pi^2 \ \Lambdabar_{ij} (\Atot)_{ji} \ 
\cr&+ \ 
2\sqrtwo \pi^2 \big( \mubar_{iu} \barhomoa_{uu} \nu_{ui}
\ - \ \nubar_{iu} \barhomoa_{uu} \mu_{ui}  \big)  }}
This is the generalization to $SU(N)$ and/or $U(N)$ of
the $SU(2)$ action presented in Eq.~(7.32) of \dkmone.

\def\Ahyp{{{\cal A}_{\rm hyp}}}
\def\hyp{{\rm hyp}}

\def\Lambdahyp{{\Lambda_\hyp}}
Next we  incorporate $N_F$ flavors of fundamental hypermultiplets.
Each such hypermultiplet comprises
a pair of $\N=1$ chiral multiplets, $Q_{f}$ and $\tilde{Q}_{f}$, with
the same conventions for component fields as in the $\N=1$ case 
discussed in Sec.~5.1.
In $\N=1$ language, these matter fields couple to the gauge multiplet
via a superpotential, 
\eqn\superp{{\rm W}\ =\ \sum_{f=1}^{N_{F}} \sqrt{2}\tilde{Q}_{f}\Phi  
Q_{f} +
m_{f}\tilde{Q}_{f}Q_{f}}
suppressing color indices. The second term is an $\N=2$ invariant mass  
term.

In what follows we will restrict our attention to the Coulomb branch of
the $\N=2$ theory,  where the hypermultiplet squarks do not acquire VEVs.
Instead, the integrations over instanton scale-sizes are regulated 
by the VEVs \vevs\ of the adjoint complex scalar $A$.
The classical component fields $\chi_f,$ $\tilde\chi_f,$ $q_f$ and $\tilde
q_f$ are still given by Eqs.~\fundredux-\squarkdredux, except that
on the Coulomb branch
the first terms on the right-hand sides of Eq.~\squarkdredux\ are 
 zero. The essential new feature in the $\N=2$ theory for $N_F>0$
is that the complex conjugate adjoint Higgs $A^\dagger$ acquires a
fermion bilinear component due the inhomogeneous term in its equation
of motion (cf.~Eq.~(5.5) in \dkmfour):
%
%
\eqn\newAdageqn{({\cal  D}^{2}\Abar)_{uv}
\  = \
{1\over{\sqrt{2}}}\,\sum_{f=1}^{N_{F}} \ \chi_{u f}
\tilde{\chi}_{f v}  \ . }
The right-hand
side comes from varying the superpotential \superp. (In contrast, the
equation for $A$ is unchanged.) 

The 
solution to Eq.~\newAdageqn\ is similar to, but simpler than, that of
\Higgseq{}. At the purely bosonic level, with all Grassmanns  
turned off, $A$ and $\Abar$ must coincide, except for
 $\vhiggs_u\rightarrow\vhiggsbar_u.$ In contrast,
the fermion bilinear contributions to 
$A$ and to $\Abar$  in the path integral are to be treated as
independent.  
This bilinear contribution to $\Abar$ is straightforwardly 
obtained from \newAdageqn,
as outlined
in Sec.~5 of \dkmfour. It has the form
\eqn\newblock{-i{\Ubar_{u'}}^{\ u+l\alpha}
\cdot
\pmatrix{0_{uv}&0 \cr
0&(\Ahyp)_{lm}\,\delta_\alpha^{\ \beta}}\cdot
U_{(v+m\beta)\, v'}}
where the $k\times k$ anti-Hermitian matrix $\Ahyp$ is defined as the
solution to the inhomogeneous linear equation
\eqn\Ahypdef{\bigL\cdot\Ahyp\ =\ \Lambdahyp }
Here the $k\times k$ anti-Hermitian matrix $\Lambdahyp$ is given by
(cf.~Eq.~(5.8) of \dkmfour): 
\eqn\Lambdahypdef{(\Lambda_\hyp)_{ij}\ =\ {i\sqrtwo\over8}\,
\sum_{f=1}^{N_F}\ \K_{if}\Kt_{fj}   }
Note that $\Lambdahyp$ and, thus, $\Ahyp$ are in fact anti-Hermitian
when it is understood that $\K^\dagger = \Kt$, that is the
Hermitian conjugation does not turn fermions into anti-fermions (see
footnote 8):
\eqn\herher{\K_{if}^\dagger = \Kt_{fi} \ , \quad
\Lambdahypdag_{ij} = - \Lambdahyp_{ji} \ , \quad
\Ahypdag_{ij} = - \Ahyp_{ji}   }


The derivation of the superinstanton
action in $\N=2$ supersymmetric $SU(N)$ and/or $U(N)$ QCD  is identical to
the one in Sec.~5 of \dkmfour; one finds
\eqn\SNFfinal{ S_{\sst \N=2 \ SQCD}^{k\hbox{-}\inst}\ =\ 
S_{\sst \N=2 \ SYM}^{k\hbox{-}\inst}\ -\
8\pi^2(\Lambdahyp)_{ij}(\Atot)_{ji}\ +\ 
\pi^{2}\sum_{f=1}^{N_{F}}m_{f}
\tilde{\cal K}_{fi}{\cal K}_{if}
}

As  with the $\N=1$ action \funactionredux, one can check that
this expression is a \susy\ invariant. On the Coulomb branch,
Eq.~\deltaKnz\ collapses to
\eqn\deltaKzero{0\ = \delta\K\ =\ \delta\Kt }
which also implies that $\Lambdahyp$ is a \susy\ invariant quantity:
\eqn\deltaLh{\delta\Lambdahyp =\ 0 }
Verifying the invariance of the action \SNFfinal\ is then a straightforward
exercise involving the transformations 
\susyalgebratwo{}, \atotr, \deltaKzero\ and \deltaLh; see Ref.~\dkmfour\
for calculational details in the $SU(2)$ case.

We should also add that the \susy\ invariance of the actions \siact\ and
\SNFfinal\ can be made more manifest, by assembling the bosonic and
fermionic collective coordinates into a space-time-constant 
$\N=2$ ``superfield,'' and reexpressing the action as an $\N=2$
``$F$-term'' constructed from this superfield; see Ref.~\dkmfour\ for
details.

\newsec{The Multi-Instanton Collective Coordinate Integration Measure}

\subsec{The overall strategy of the construction}

In general, semiclassical physics requires the answers to two types
of questions: \hbox{(1) What} are the configurations that dominate the path
integral? and (2) How does one properly weight these configurations?
For instanton processes in supersymmetric theories, we have given
the answer to (1) in Secs.~2-3 above, by  constructing 
the $\N=1$ and $\N=2$
ADHM supermultiplets. In this section we construct the corresponding
 collective coordinate integration measures for all topological
numbers $k$, thereby answering
question (2).
The cluster decomposition property of the
measure is checked in the Appendix.
 Further calculational details may be found in Ref.~\dkmten,
for the gauge group $SU(2)$. 

 As the small-fluctuations determinants in a self-dual
background cancel between the bosonic and fermionic sectors in a
supersymmetric theory \dadda, 
the relevant measure is the one inherited from the 
 path integral on changing variables from the fields 
to the collective coordinates which parametrize the instanton moduli space
${\rm M}^k_{\rm phys}$. In principle the super-Jacobian for this change of
variables can be calculated by evaluating the normalization matrices of
the appropriate bosonic and fermionic zero-modes. 
In practice, this involves solving the nonlinear
ADHM constraints \fcone a and can
only be accomplished for $k\leq 2$ \Osborn.

Following our earlier work \dkmten,  we
pursue an alternative approach to the problem of determining the
correct measure. Rather than attempt to implement the bosonic and
fermionic constraints, which cannot be done explicitly for $k>3,$ 
the measure will be expressed in terms of the original
overcomplete, unconstrained matrices of collective coordinates. The
requisite constraints will be introduced, by hand, as $\delta$-functions
in the integrand. (An analogy would be the measure
$dx\,dy\,\delta(x^2+y^2-1)$ rather than  $d\theta$ for integration
on a circle.) As with the action, we will demand that the resulting
measure be a \susic\ invariant quantity. The reason our construction
can work is that the various bosonic and fermionic constraints actually
form a supermultiplet of constraints, as mentioned in Sec.~3.
See Ref.~\dkmfour\ for further discussion of this point.

In Sec.~8, we will give a concrete example of the usefulness of our measure.
We will see that the $\delta$-function constraints are best exponentiated
through the introduction of a supermultiplet of Lagrange multipliers,
after which the original collective coordinates in the problem can be
entirely integrated out (the exponent is Gaussian in these variables).
What is left is an effective measure for the Lagrange multipliers 
 dual to the constraints. The resulting integration over these new
variables can then be carried out by a suitable application of Stoke's
theorem.

The first step in the construction of the measure
is to formally undo the $U(k)$ quotient
described in Eq.~\modspace\ and define an unidentified measure, $\dmuk$, for
integration over the larger moduli space ${\rm M}^k\,$:
\eqn\dmudefI{\int_{{\rm M}^k_{\rm phys}}\dmuphys\ =\
{1\over\VolUk}\,\int_{{\rm M}^k}\dmuk}
The correctly normalized volumes for the $U(k)$ groups 
\eqn\voluk{
{\rm Vol}\big(U(k) \big) \ = \ 
2^{2k-1} \pi^{k^2+2k-1 \over 2}
\ \prod_{i=1}^{k-1}
\frac{1}{\Gamma(i+\hf)} 
}
follow from
\eqn\volgrr{
{U(k) \over U(k-1) \times U(1)} \ = \ S^{2(k-1)}}
together with the initial condition
\eqn\initcond{{\rm Vol}\big(U(1)\big)\ =\ 2\pi\ . }
$S^{2(k-1)}$ is the $2(k-1)$-sphere and 
\eqn\volsph{
{\rm Vol}\big(S^{2(k-1)} \big) \ = \ 
\frac{2\pi^{k-\hf}}{\Gamma(k-\hf)}\ . }
In addition to being a \susic\ invariant, we will demand that the measure
transform as a singlet under this residual $U(k)$.

We  now propose explicit expressions for 
 $\dmuk$ in both the $\N=1$ and $\N=2$ cases. (A similar construction
works for the $\N=4$ case as well, while for the nonsupersymmetric
case complications arise due to the reemergence of the
small-fluctuations determinants \dkmten.) We will then argue that
these proposals are in fact  unique. 

\subsec{The case of $\N=1$ \susy}

Following the strategy outlined above, we can immediately write down
the ansatz for the $\N=1$ \susic\ collective coordinate
integration measure in topological \hbox{sector $k\,$:}
\eqn\dmudef{\eqalign{\int\dmuphys\ &\equiv\ {1\over\VolUk}\,\int\dmuk
\cr&=\ {C_1^k\over\VolUk}
\int d^{2Nk} \wbar \ d^{2Nk} w \ d^{Nk} \mubar \ d^{Nk} \mu 
\ d^{4k^2} a' \ d^{2k^2} \M'
\cr&\times\ 
\big[\prod_{c=1}^3 
\delta^{(k^2)}\big(\trtwo ( \hf\tau^c \abar a) \big)
 \big]
\ \delta^{(2k^2)} \big(\Mbar a + \abar \M \big)
}}
The differentials in Eq.~\dmudef\ have the following
explicit meanings:
\eqna\aint
$$\eqalignno{
\int d^{4k^2} a' \ &= \ 
 \int \prod_{m=0}^3 \big[ \prod_{i=1}^k d a^{\prime m}_{ ii} \big]
\big[ \prod_{1\le i < j\le k} d \ {\rm Re}(a^{\prime m}_{ ij}) \ 
d \ {\rm Im}(a^{\prime m}_{ ij}) \big]
&\aint a\cr
\int d^{2Nk} \wbar \ d^{2Nk} w 
 \ &= \ 
\int \prod_{\dalpha=1,2} \prod_{u=1}^N \prod_{i=1}^k
d \wbar^\dalpha_{iu} \ dw_{ui\dalpha} 
&\aint b\cr
\int d^{2k^2} \M'\ &=\ 
\int  \prod_{\alpha=1,2}
\big[ \prod_{i=1}^k d \M'_{\alpha ii} \big]
\big[ \prod_{1\le i < j\le k} d \ {\rm Re}(\M'_{\alpha ij}) \ 
d \ {\rm Im}(\M'_{\alpha ij}) \big]
&\aint c\cr
\int  d^{Nk} \mubar \ d^{Nk} \mu 
\ &=\ 
\int  \prod_{u=1}^N \prod_{i=1}^k d\mubar_{iu} \ d\mu_{ui} 
&\aint d\cr}$$
Notice that these expressions presuppose
the canonical form \aaa{} for $b$, so that
the collective coordinate matrices $a$ and $\M$ are assumed from the
outset to
satisfy Eqs.~\fcone b and \mctw, respectively. The remaining constraints,
namely \fcone a and \zmcon a, are implemented explicitly in Eq.~\dmudef\
via the $\delta$-functions. These have the explicit meanings:
\eqna\bcns
$$\eqalignno{\prod_{c=1}^3 \ 
\delta^{(k^2)}\big(\trtwo (\hf \tau^c \abar a) \big)
\ &= \ \prod_{c=1}^3 \big[\prod_{i=1}^k 
\delta\big(\trtwo (\hf \tau^c \abar a)_{ii} \big) \big]   
&\bcns a\cr
&\times\ \big[ \prod_{1\le i < j\le k} 
\delta\big(\trtwo {\rm Re}( \hf\tau^c \abar a)_{ij} \big)
\ \delta\big(\trtwo {\rm Im}(\hf \tau^c \abar a)_{ij} \big)
\big]  &{}\cr
\delta^{(2k^2)} \big(\Mbar a + \abar \M \big)\ &=\
\prod_{\dalpha=1,2} \bigl[\prod_{i=1}^k 
\delta\big(\Mbar a_\dalpha + \abar_\dalpha \M
\big)_{ii} \big]   
&\bcns b\cr
&\times\ \big[ \prod_{1\le i < j\le k} 
\delta\big(\ {\rm Re}\big(\Mbar a_\dalpha + 
\abar_\dalpha \M
\big)\big)_{ij} \ 
\delta\big( \ {\rm Im}\big(\Mbar a_\dalpha + 
\abar_\dalpha \M\big)
\big)_{ij} 
\big] 
&{}\cr}$$

%
%
%
%
%
%
%

We can make the following arguments in support of the proposed measure
\dmudef:

(i) In the $1$-instanton sector, Eq.~\dmudef\ reduces to
\eqn\onesimp{\eqalign{\int\dmuonephys\ 
\ &=\  {C_1 \over 2 \pi}
\int d^{4} a' \ d^{2} \M'\ d^{2N} \wbar \ d^{2N} w \ 
d^{N} \mubar \ d^{N} \mu 
\cr&\times\ \big[\prod_{c=1}^3 \delta\big(\trtwo
 (\hf \tau^c \wbar w) \big)\big]
\ \delta^2 \big(\mubar w + \wbar \mu \big)
}}
After one resolves the $\delta$-function constraints as per \omom{}, this 
precisely reproduces 
 the standard 't Hooft-Bernard 1-instanton measure \refs{\tHooft,\Bernard}. 
In particular the position $X_m$, 
size $\rho$, and group
iso-orientation $\Omega$ of the instanton can be deduced, respectively,
from
Eqs.~\xcon, \wcon\ and \omom{}. Likewise, the fermionic collective coordinates
in Eq.~\onesimp\  can be identified with the ``supersymmetric'' and
``superconformal'' modes \susymo-\suconmo, and with the superpartners
of the iso-orientations zero modes discussed at the end of Sec.~3.1
 Also $C_1$ is a scheme-dependent 1-instanton factor
which can be derived from the pure Yang-Mills 1-instanton factor
in \Bernard. It is pleasing that the ADHM parametrization replaces
the trigonometric variables of the coset matrix $\Omega$ describing
the embedding of the instanton into $SU(N)$, by Cartesian variables
endowed with a flat measure (apart from the $\delta$-function insertions).

(ii) The power of length carried by the $k$-instanton measure should be
$b_0 k=3Nk$. 
Since $[a]=1$, $[\mu]=1/2$ and $[d \mu]=-1/2$,
the right-hand side of
Eq.~\dmudef\ does have the correct engineering dimension.

(iii) The anomalous $U(1)_{R}$ symmetry requires
 a net  of $2Nk$ unsaturated Grassmann
integrations in the $k$-instanton measure, 
in other words, $2Nk$ exact fermion zero modes. 
It is easy to see that this counting is obeyed by
the right-hand side of Eq.~\dmudef: $2k^2$ fermionic $\delta$-functions
saturate $2k^2$ out of the $2k^2+2Nk$ fermionic integrations over
$\M'$, $\mubar$ and $\mu$ leaving $2Nk$ exact fermion zero modes.

(iv) The $U(k)$ invariance of the measure \dmudef\ is obvious.

(v) Cluster decomposition in the dilute-gas limit  of large
space-time separation between instantons fixes the overall constant 
in the $k$-instanton measure \dmudef\ in terms of the $1$-instanton
factor $C_1$. The derivation is analogous to that in  
\dkmten\ and is detailed in the Appendix.

(vi) As discussed above, the $k$-instanton measure has to be a \susic\
 invariant.
This important requirement can be directly checked 
performing the \susy\ transformations \susyalgebra{}
in the integrand  of  Eq.~\dmudef.
For $-i\xi Q$ the first $\delta$-function 
in \dmudef\ is trivially invariant while the argument of the second
$\delta$-function also does not change due to \cone b. For $i\xibar\Qbar$
the reasoning is as follows: the argument of the
second $\delta$-function in \dmudef\ 
is invariant, while that of the
first  $\delta$-function 
transforms into itself plus an admixture of  the second, 
so that the product of 
$\delta$-functions is an invariant. 

(vii) Finally we can make the following uniqueness argument \dkmten.
Since the \deltafcns\ in Eq.~\dmudef\ are dictated by the ADHM formalism,
and since the resulting measure turns out to be
 a \susy\ invariant and also has the correct transformation property under
the anomalous $U(1)_{R}$ symmetry,
we claim that the ansatz \dmudef\ is in fact unique.
To see why, let us consider including an additional
function of the collective coordinates, $f(a,\M),$ in the integrand
of Eq.~\dmudef. To preserve \susy, we  require that $f$
be a \susy\ invariant.  It is a fact that 
any non-constant function that is a \susy\ invariant
must contain fermion bilinear pieces (and possibly higher powers of
fermions as well). By the rules of Grassmann integration, such bilinears
would necessarily
lift some of the adjoint fermion zero modes contained in $\M.$ But since
Eq.~\dmudef\ contains precisely the right number
of unlifted fermion zero modes dictated by the $U(1)_{R}$ anomaly, 
namely $2Nk$, this argument rules
out the existence of a non-constant function $f$. Moreover, any
constant $f$ would be absorbed into the overall
multiplicative factor, which is to be fixed by cluster decomposition. 
A similar uniqueness argument applies
to our proposed
ADHM measure for $N=2$ theories discussed below.

The measure \dmudef\ is easily augmented to incorporate fundamental
matter multiplets, as discussed in Sec.~5.1. Since the Higgsinos
satisfy the normalization condition \CGTone
\eqn\orth{\int\, d^{4}x \ 
\tilde{\chi}^{}_{f\alpha u}\, \chi^{\alpha}_{u f'}
\ = \ 
\pi^{2} \Kt_{fi} \K_{if'}\ ,}
the normalized hypermultiplet part of the 
$k$-instanton measure  reads
\dkmfour
\eqn\muhypd{\int d\muhyp^{(k)}\ =\
{1\over\pi^{2kN_F}}\int\prod_{f=1}^{N_F}d\K_{1f}\cdots d\K_{kf}\,
d\Kt_{f1}\cdots d\Kt_{fk} \ . }
The total measure is then simply the product
\eqn\totmeason{\dmuphys\times d\muhyp^{(k)} }

\subsec{The case of $\N=2$ \susy}

Next we turn to the $\N=2$ measure. We can be especially brief, as
the construction is an obvious extension of Eq.~\dmudef. The new features
are the presence of the second adjoint fermion $\psi$ described by the
collective coordinate matrix $\N$, and also, the adjoint Higgs whose
associated collective coordinate $\Atot$ is subject to the
``spin-0'' constraint \thirtysomething. As before, we postulate:
%
\eqn\dmutwodef{\eqalign{\int \dmuphys\ &\equiv\ {1\over\VolUk}\,\int\dmuk
\cr&=\ {(C'_1)^k\over\VolUk}
\int d^{2Nk} \wbar \ d^{2Nk} w \ d^{Nk} \mubar \ d^{Nk} \mu 
\ d^{Nk} \nubar \ d^{Nk} \nu 
\cr&\times\ d^{4k^2} a' \ d^{2k^2} \M'  \ d^{2k^2} \N' \ d^{k^2} \Atot
\cr&\times\ 
\big[\prod_{c=1}^3 
\delta^{(k^2)}\big(\trtwo (\hf \tau^c \abar a) \big)
 \big]
\ \delta^{(2k^2)} \big(\Mbar a + \abar \M \big)
\ \delta^{(2k^2)} \big(\Nbar a + \abar \N \big)
\cr&\times \ 
\delta^{(k^2)}\big( \bigL\cdot\Atot - \Lambda - \Lambda_f \big)
}}

The arguments (i)-(vii) of Sec.~ 6.2 can again be made for this $\N=2$ measure,
with the obvious modifications that now there are twice as many
adjoint
fermionic zero modes dictated by the anomaly, and also the \susy\ algebra
is enhanced to incorporate Eqs.~\susyalgebratwo{} and \atotr. In the
presence of $N_F$ matter hypermultiplets, the measure is again enlarged
to the form \totmeason.

\newsec{Explicit Expression for the $\N=2$ Prepotential}

We now discuss the prepotential $\F(A)$, whose derivatives govern
the Wilsonian effective action of the theory, as follows:
\eqn\WilsF{{\cal L}_{\rm eff}\ =\ 
{\rm Im}\,{1\over4\pi}\,\Big[\,
\int d^4\theta\,{\partial\F(A)\over\partial A_u}\,\bar A_u\ +\ 
{1\over2}\,\int d^2\theta\,{\partial^2\F(A)\over\partial A_u\,\partial A_v}
\,W_u\,W_v\,\Big]\ .}
Here $A_u$ and $W_u$ are the $\N=1$ chiral superfields containing
the $u$th massless Higgs boson and the $u$th photon field strength,
respectively, after the spontaneous gauge symmetry
breakdown $SU(N)\rightarrow U(1)^{N-1}$ 
or $U(N)\rightarrow U(1)^{N}$ 
induced by the Higgs VEVs \vevs.

In earlier work \refs{\dkmfour,\dkmten,\dkmtwo} 
we presented a general formula for the
$k$-instanton contribution to the prepotential of the $\N=2$
supersymmetric QCD. With $\F_k$ defined as in Eq.~\multiexp, we derived:
\eqn\Fnfinal{{\cal F}_k
\ =\ 8\pi i\int\dmuphystilde\,\exp\big[
-S_{\sst \N=2 \ SQCD}^{k\hbox{-}\inst}\big]}
Here $\dmuphystilde$ is the ``reduced measure''
which is obtained from the physical $\N=2$ measure,
$\dmuphys$, as follows:
\eqn\mufactor{\int\dmuphys\ =\ \int d^4x_0\,d^2\xi_1\,d^2\xi_2\,
\int\dmuphystilde}
where $(x_0,\xi_1,\xi_2)$ gives the global position of the 
multi-instanton in $\N=2$ superspace. Explicitly, $x_0,$ $\xi_1$ and
$\xi_2$ are the linear combinations proportional to the `trace' components
of the $k\times k$ matrices $a',$ $\M'$ and $\N'$, respectively \dkmone:
\eqn\traceparts{x_0\ =\ {1\over k}\Tr_k\,a'\ ,\quad
\xi_1\ =\ {1\over4k}\Tr_k\M'\ ,\quad
\xi_2\ =\ {1\over4k}\Tr_k\N'\ .}
Note that these $\N=2$ superspace modes do not enter into the
$\delta$-function constraints in \dmutwodef\
and so do indeed factor out in this simple way. Furthermore, the four
exact \susic\ modes $\xi_{1\alpha}$ and 
$\xi_{2\alpha}$ are the only fermionic modes that are not lifted by
(i.e., do not appear in) the action \siact, \SNFfinal.

Given these expressions for the prepotential, one also knows the
all-instanton-orders expansion of the quantum modulus
$u_2=\langle\Tr\,A^2\rangle$, since on general grounds 
\eqn\matrelis{
u_2(\vhiggs_1,\ldots,\vhiggs_N){{}\atop\Big|}_{k\hbox{-}{\rm inst}}\ 
=\ 2i\pi k\cdot
\F_{k}(\vhiggs_1,\ldots,\vhiggs_N)
}
This relation was originally derived by Matone
\Matone\ for the gauge group $SU(2)$, but the all-instanton-orders  proof 
of it presented in Refs.~\refs{\dkmfour,\dkmtwo} is valid for the general cases
$SU(N)$ and/or $U(N)$, as the reader can verify (see also \Fucito).

The above collective coordinate integral expression for $\F_k$
constitutes
a closed series solution, in quadratures, of the low-energy dynamics of the
Coulomb branches of the
$\N=2$ models. It is noteworthy that this solution
is obtained purely from the semiclassical regime, without appeal
to duality.

\newsec{One-instanton Contribution to the Prepotential}
\def\abar{\bar{\rm v}}

In this section we explicitly evaluate the 1-instanton contribution
to the prepotential, $\F_1,$ starting with the
integral expression \Fnfinal. From the formulae for the $\N=2$
action and reduced measure given respectively by Eq.~\SNFfinal\ and  by
Eqs.~\dmutwodef\ and \mufactor, one writes down:
\eqn\integraldef{\eqalign{\Foneinst\ &=\ 
{iC_1'\over2^6\pi^{2N_F}}\,\int d\Atot\cdot
\prod_{u=1}^Nd\mubar_ud\mu_u d\nubar_ud\nu_ud^2\wbar_u^\dalpha
d^2w_{u\dbeta}\cdot\prod_{f=1}^{N_F}d\K_fd\Kt_f
\cr&\times\ 
\delta\big(\bigL\cdot\Atot-\Lambdatot\big)
\prod_{c=1,2,3}\delta\big(\hf(\tau^c)^\dalpha_{\ \dbeta}\,\wbar_u^\dbeta
w^{}_{u\dalpha}\big)
\cr&\times\ 
\prod_{\dalpha=1,2}\delta\big(\mubar_uw_{u\dalpha}+\wbar_{u\dalpha}
\mu_u\big)\delta\big(\nubar_uw_{u\dalpha}+\wbar_{u\dalpha}
\nu_u\big)
\cr&\times\ 
\exp\Big(\,-8\pi^2|{\rm v}_u|^2\wbar_u^\dalpha w^{}_{u\dalpha}+2\sqrtwo\pi^2i
(\mubar_u\abar_u\nu_u-\nubar_u\abar_u\mu_u)
\cr&\qquad\quad+\ 8\pi^2(\Lambdabar+\Lambdahyp)\Atot
-\pi^2\sum_{f=1}^{N_F}m_f\Kt_f\K_f\ \Big)
}}
Here the scheme-dependent 1-instanton factor $C_1'$, from Eq.~\dmutwodef,
is proportional to $\Lambda^{2N-N_F}$ where $\Lambda$ is the dynamically
generated scale.

To evaluate this integral, it is helpful to exponentiate the various
$\delta$-functions by means of Lagrange multipliers, and to interchange
the resulting order of integration. In other words, one integrates
out the ADHM supermultiplet $\{a,\M,\N,\Atot\}$ first, and next
the hypermultiplet collective coordinates $\K_f$ and $\Kt_f$, 
and only then performs
the integration over the Lagrange multipliers.

The spin-1 and spin-$1/2$
constraints in Eq.~\integraldef\
are exponentiated in the usual manner, respectively as:
\def\bp{{\bf p}}
\eqn\spinone{\prod_{c=1,2,3}
\delta\big(\hf(\tau^c)^\dalpha_{\ \dbeta}\,\wbar_u^\dbeta
w^{}_{u\dalpha}\big)\ =\ {1\over\pi^3}\,\int d^3\bp\,
\exp(ip^c(\tau^c)^\dalpha_{\ \dbeta}\,\wbar_u^\dbeta w^{}_{u\dalpha})\ ,}
and
\eqna\spinhf
$$\eqalignno{\prod_{\dalpha=1,2}
\delta\big(\mubar_uw_{u\dalpha}+\wbar_{u\dalpha}\mu_u\big)
\ &=\ 2\int d^2\xi\,\exp\big(\xi^\dalpha(\mubar_uw_{u\dalpha}
+\wbar_{u\dalpha}\mu_u)\big)
&\spinhf a
\cr
\prod_{\dalpha=1,2}
\delta\big(\nubar_uw_{u\dalpha}+\wbar_{u\dalpha}\nu_u\big)
\ &=\ 2\int d^2\eta\,\exp\big(\eta^\dalpha(\nubar_uw_{u\dalpha}
+\wbar_{u\dalpha}\nu_u)\big)
&\spinhf b}$$
In this way we introduce the triplet of bosonic Lagrange multipliers
$p^c$, as well as the Grassmann spinor Lagrange multipliers
$\xi^\dalpha$ and $\eta^\dalpha.$ 
The exponentiation of the spin-0 constraint is best accomplished in a slightly
trickier way involving a term in the action, as follows:
\def\Re{{\rm Re\,}}
\def\zbar{{\bar z}}
\def\Im{{\rm Im\,}}
\eqn\spinzero{\eqalign{&\int 
d\Atot\,\delta\big(\bigL\cdot\Atot-\Lambdatot\big)
\exp\big(8\pi^2(\Lambdabar+\Lambdahyp)\Atot\big)\cr
&=\ {1\over\det\bigL}\,
\exp\big(8\pi^2(\Lambdabar+\Lambdahyp)\cdot\bigL^{-1}\cdot\Lambdatot\big)
\cr
&=\ 8\pi\int d(\Re z)d(\Im z)\,\exp\big(-8\pi^2(\zbar\,\bigL\, z-
(\Lambdabar+\Lambdahyp)z-\zbar\Lambdatot)\big)}}
The  second equality  follows from the general Gaussian identity
\eqn\gaussid{\int\prod_i
d(\Re z_i)d(\Im z_i)\,\exp\big(-\zbar_iK_{ij}z_j+\bar y_iz_i+\zbar_iy_i\big)
\ =\ {1\over\det(K/\pi)}\,\exp(\bar y_iK^{-1}_{ij}y_j)}
which can be used to exponentiate the spin-0 constraint in an elegant
way for arbitrary instanton number $k$. 
The advantage of the rewrite \spinzero\ is that $\bigL$ is easier to
manipulate in the exponent than $\bigL^{-1}$ (which appears implicitly 
in the definition of $\Atot$).
In the present case, with $k=1,$
the operator $\bigL$ collapses to a $1\times1$ $c$-number matrix:
\eqn\bigLcollapse{\bigL\ =\ \det\bigL\ =\ \wbar_u^\dalpha w^{}_{u\dalpha}\ .}
Likewise $\Lambdabar$ and $\Lambdatot$ collapse to
\eqn\alsocollapse{\Lambdabar=-i\abar_u\wbar_u^\dalpha w^{}_{u\dalpha}\ ,
\qquad \Lambdatot=i{\rm v}_u\wbar_u^\dalpha w^{}_{u\dalpha}-{1\over2\sqrtwo}
(\nubar_u\mu_u-\mubar_u\nu_u)\ .}

\def\alphabar{\bar\alpha}
Now consider the combined exponent formed from Eqs.~\integraldef-\spinzero.
The linear shifts
\eqn\linshift{\eqalign{\mu_u&\rightarrow\mu_u+{i\eta^\dalpha w_{u\dalpha}\over
2\sqrtwo\pi^2\alphabar_u},\quad
\mubar_u\rightarrow\mubar_u+{i\eta^\dalpha \wbar_{u\dalpha}\over
2\sqrtwo\pi^2\alphabar_u},\quad\cr
\nu_u&\rightarrow\nu_u-{i\xi^\dalpha w_{u\dalpha}\over
2\sqrtwo\pi^2\alphabar_u},\quad
\nubar_u\rightarrow\nubar_u-{i\xi^\dalpha \wbar_{u\dalpha}\over
2\sqrtwo\pi^2\alphabar_u}}}
eliminate the linear terms in these variables.
By inspection, the Grassmann
integrations over  $\{\mu_u,\nu_u,\mubar_u,\nubar_u\}$ 
then simply bring down a factor of 
\eqn\prodown{\prod_{u=1}^N(2\sqrtwo\pi^2i\alphabar_u)^2}
In Eqs.~\linshift-\prodown, we have
 defined $\alpha_u$ and $\alphabar_u$ as the naturally appearing
linear combinations
\eqn\lincomb{\alpha_u\ =\ {\rm v}_u+iz\ ,\qquad\alphabar_u\ =\ \abar_u-i\zbar\ .}
Next, the $\{w_u,\wbar_u,\K_f,\Kt_f\}$ integrations are accomplished, using the
identities
\eqn\wintid{\int d^2w_ud^2\wbar_u\,\exp\big(-A^0\wbar_u^\dalpha
w^{}_{u\dalpha}+i\sum_{c=1,2,3}A^c(\tau^c)^\dalpha_{\ \dbeta}\,\wbar_u^\dbeta
w^{}_{u\dalpha}\big)\ =\ {-4\pi^2\over (A^0)^2+\sum(A^c)^2}\ }
and
\eqn\KKtint{\int\prod_{f=1}^{N_F}d\K_fd\Kt_f\,
\exp\Big(\,8\pi^2\Lambdahyp\,z
-\pi^2\sum_{f=1}^{N_F}m_f\Kt_f\K_f\ \Big)\ =\
\pi^{2N_F}\,\prod_{f=1}^{N_F}(i\sqrtwo z+m_f)\ .}
In this way, all the original
ADHM variables $\{a,\M,\N,\Atot,\K,\Kt\}$ are eliminated
from the integral \integraldef. One is left with an integral
over Lagrange multipliers only:
\def\calB{{\cal B}}
\eqn\newintegral{\Foneinst\ =\ 
{iC_1'\over2\pi^2}\,\int d^3\bp\, d^2\xi d^2\eta
\,d(\Re z)d(\Im z)\,\calB\prod_{f=1}^{N_F}(i\sqrtwo z+m_f)}
where
\eqn\calBdef{\calB\ =\ 
\prod_{u=1}^N\ {(2\sqrtwo\pi^2 i\alphabar_u)^2(-4\pi^2)\over
\big(8\pi^2|\alpha_u|^2\big)^2+\sum_{c=1,2,3}\,(p^c+\Xi^c_u)^2}}
and $\Xi^c_u$ is the fermion bilinear
\eqn\Xidef{\Xi^c_u\ =\ {1\over4\sqrtwo\pi^2\alphabar_u}
\big(\xi^{}_\dalpha(\tau^c)^\dalpha_{\ \dbeta}\,\eta^\dbeta-
\eta^{}_\dalpha(\tau^c)^\dalpha_{\ \dbeta}\,\xi^\dbeta\big)\ .}
When $N_F=0$ the product over flavors in Eq.~\newintegral\ should simply
be replaced by unity.

The $\{\xi,\eta\}$ Grassmann integrations in Eq.~\newintegral\
must be saturated with two insertions of $\Xi$:
\eqn\Xiinsert{\int d^2\xi d^2\eta\,\Xi^b_u\,\Xi^c_v\ =\ 
{\delta^{bc}\over16\pi^4\alphabar_u\alphabar_v}\ .}
Extracting these quadratic powers of $\Xi$ from $\calB$ can be done
quite elegantly, thanks to the algebraic identity
\eqn\niceid{\eqalign{\int d^2\xi d^2\eta\,\calB\ &=\ 
\sum_{b,c=1}^3\,\sum_{u,v=1}^N{\delta^{bc}\over16\pi^4\alphabar_u\alphabar_v}
\cdot{1\over2}{\partial^2\over\partial\Xi^b_u\,\partial\Xi^c_v}\,\calB
{\Big|}_{\Xi=0}
\cr&=\ {1\over32\pi^4|\bp|^2}\,\Big(\sum_{u=1}^N{\partial\over\partial
\abar_u}\Big)^2\,\calB{\Big|}_{\Xi=0}\ .}}
Pulling the VEV derivatives outside the integral, one therefore finds:
\eqn\integralb{\Foneinst\ =\ 
{iC_1'\over2\pi^2}\cdot{1\over32\pi^4}
\Big(\sum_{u=1}^N{\partial\over\partial\abar_u}\Big)^2
\,\int d(\Re z)d(\Im z)\,\Gamma\,\prod_{f=1}^{N_F}(i\sqrtwo z+m_f)\ .}
Here
\eqn\Gammadef{\Gamma\ =\ \int d^3\bp\,{1\over|\bp|^2}\,
\prod_{u=1}^N\ {(2\sqrtwo\pi^2 i\alphabar_u)^2(-4\pi^2)\over
\big(8\pi^2|\alpha_u|^2\big)^2+|\bp|^2}\
=\ 8\pi^6\sum_{u=1}^N\,{\alphabar_u\over\alpha_u}\prod_{v\neq u}\,
{\pi^2\over2}{\alphabar^2_v\over|\alpha_v|^4-|\alpha_u|^4}\ ,}
the second equality following from a standard contour integration in
the variable $|\bp|$, extended to run from $-\infty$ to $\infty.$

In this fashion, the original expression \integraldef\ has collapsed
to a 2-dimensional integral over the $xy$ plane (with $x=\Re z$ and
$y=\Im z$ henceforth). To complete this integration, notice that the only
dependence on $\abar_u$ in the integrand is through the variables
$\alphabar_u=\abar_u-i\zbar$. Therefore, it is tempting---but incorrect---to
pull the $\abar_u$ derivatives back inside the integrand, and to
make the naive replacement
\eqn\replace{\sum_{u=1}^N{\partial\over\partial\abar_u}\ \rightarrow\
i\,{\partial\over\partial\zbar}\ ,\qquad
\Big(\sum_{u=1}^N{\partial\over\partial\abar_u}\Big)^2\ \rightarrow\
-\Big({\partial\over\partial\zbar}\Big)^2\ .}
\eqna\exampole
The error here is due to the fact that the two sides of Eq.~\replace\ can
differ by $\delta$-function contributions which arise at the locations
of poles in the $z$ variable. As a simple example, whereas obviously
$\big(\sum\,\partial/\partial\abar\,\big)\,z^{-1}=0,$ one also has,
in contrast,\foot{The normalization factor on the right-hand side
of Eq.~\exampole a is easily fixed by integrating both sides
against $\exp(-\lambda z\zbar)$.}
$$\eqalignno{
{\partial\over\partial\zbar}\,{1\over z}\ &=\ \pi\, \delta(x)\delta(y)
\ ,&\exampole a
\cr
\Big({\partial\over\partial\zbar}\Big)^2\,{1\over z}\ &=\ 
\pi\,{\partial\over\partial\zbar}\, \delta(x)\delta(y)\ =\ 
{\pi\over2}\,\big(
\delta'(x)\delta(y)+i\delta(x)\delta'(y)\big)\ .&\exampole b}$$
The lesson is that one can legitimately
trade $\abar_u$ differentiation for $\zbar$
differentiation as per Eq.~\replace---but only if one explicitly
subtracts off the
extraneous $\delta$-function pieces that are generated at
the locations of the poles in $z$. Accordingly, we can split up 
$\Foneinst$ into two parts,
\def\calF{{\cal F}}
\eqn\Fsplit{\Foneinst\ =\ \calF_\delta\ +\ \calF_\partial\ ,}
where $\calF_\delta$ 
is the contribution of these $\delta$-function corrections,
while $\calF_\partial$ is a boundary term arising from judicious use
of Stoke's theorem applied to $\partial^2/\partial\zbar^2.$ Let us evaluate
each of these parts, in turn:

\def\bigR{{\rm I}\!{\rm R}}
\leftline{\ \ \ \it Calculation of $\calF_\delta$.}

As stated, to calculate $\calF_\delta,$ one converts $(\sum\partial/\partial
\abar_u)^2$ into $-\partial^2/\partial\zbar^2$ as per Eq.~\replace,
then subtracts off the spurious $\delta$-function contributions that
correspond to the poles in $z$ of the expression $\Gamma$ given in
Eq.~\Gammadef. The relevant poles lie at the $N$ distinct values
\eqn\poledef{0\ =\ \alpha_u\ = {\rm v}_u+iz\ =\ (\Re {\rm v}_u-y)+i(\Im {\rm v}_u+x)\ 
.}
There also appear to be poles in $\Gamma$ when $|\alpha_v|^2=
\pm |\alpha_u|^2$ but these are irrelevant: the poles at 
$|\alpha_v|^2=-|\alpha_u|^2$ lie away from the real domain of
integration  $(x,y)\,\in\,\bigR^2,$ whereas the poles at
$|\alpha_v|^2=+|\alpha_u|^2$ have residues that cancel pairwise
among the terms in Eq.~\Gammadef\ (these pairs correspond to
interchanging the indices $u$ and $v$). Restricting our attention to
the singularities \poledef, we therefore find:
\eqn\deltastuff{\eqalign{\calF_\delta\ &=\ 
{iC_1'\over2\pi^2}\cdot{1\over32\pi^4}\cdot8\pi^6\,
\int dx\,dy\,
\sum_{u=1}^N\,\Big[\,\Big({\partial^2\over\partial\zbar^2}\,{1\over\alpha_u}
\Big)\ +\ 2\Big({\partial\over\partial\zbar}\,{1\over\alpha_u}\Big)
{\partial\over\partial\zbar}\,\Big]
\cr &\times\ \Big[\,\alphabar_u\,
\prod_{v\neq u}\,
{\pi^2\over2}{\alphabar^2_v\over|\alpha_v|^4-|\alpha_u|^4}\,\Big]
\,\prod_{f=1}^{N_F}(i\sqrtwo z+m_f)\ .}}
Integrating the first term on the right-hand side (the $\partial^2/\partial
\zbar^2$ term) once by parts  cancels half the second term, whereupon the
identity
\eqn\alphaid{{\partial\over\partial\zbar}\,{1\over\alpha_u}\ =\ 
-i\pi\,\delta(\Im {\rm v}_u+x)\delta(\Re {\rm v}_u-y)}
[cf.~Eqs.~\exampole a and \poledef] quickly leads to
\eqn\Fdelfinal{\calF_\delta\ =\ -{iC_1'\pi^{2N-1}\over2^{N+2}}
\,\sum_{u=1}^N\,\prod_{v\neq u}\,{1\over({\rm v}_v-{\rm v}_u)^2}\,
\prod_{f=1}^{N_F}(-\sqrtwo {\rm v}_u+m_f)\ .}

\vfil\eject
\leftline{\ \ \ \it Calculation of $\calF_\partial$.}

\def\calD{{\cal D}}
Next we evaluate the boundary term $\calF_\partial$ implied by the
naive replacement \replace. It is useful to switch to polar coordinates,
$(x,y)\rightarrow(r,\theta),$ in terms of which
\eqn\zbarderiv{{\partial^2\over\partial\zbar^2}\ =\ 
{1\over r}\,{\partial\over\partial r}\circ\calD_r\ +\
{\partial\over\partial\theta}\circ\calD_\theta}
where
\eqn\calDdef{\calD_r\ =\ \fourth e^{2i\theta}\,\big(\,
2+r\,{\partial\over\partial r}\,\big)\ ,\qquad
\calD_\theta\ =\ {i\over4r^2}\,e^{2i\theta}\,\big(\,
1+2r{\partial\over\partial r}
+i\,{\partial\over\partial\theta}\,\big)\ .}
Since the integrand in Eq.~\integralb\ is a single-valued function
of $\theta,$ the $(\partial/\partial\theta)\,
\calD_\theta$ term can be neglected.
Stoke's theorem then equates the 2-dimensional integral \integralb\
to the angularly integrated action of $\calD_r$ evaluated on the circle
of infinitely large radius:
\eqn\Fddef{\eqalign{\calF_\partial\ &=\
-{iC_1'\over2\pi^2}\cdot{1\over32\pi^4}\cdot 8\pi^6\,
\lim_{r\rightarrow\infty}\,\quarter\big(\,2+r\,{\partial\over\partial r}\,\big)
\cr&\times\ \int_0^\infty d\theta\,e^{2i\theta}\,
\Big[\,\sum_{u=1}^N\,{\alphabar_u\over\alpha_u}\prod_{v\neq u}\,
{\pi^2\over2}{\alphabar^2_v\over|\alpha_v|^4-|\alpha_u|^4}\,\Big]
\,\prod_{f=1}^{N_F}(i\sqrtwo re^{i\theta}+m_f)\ ,}}
where $\alpha_u={\rm v}_u+ire^{i\theta}$ and $\alphabar_u=\abar_u-ire^{-i\theta}$.
The remaining $\theta$ integral is best evaluated by changing variables to
$\xi=e^{i\theta}$, and summing the poles in $\xi$ which sit within the
unit circle. These lie at the points where $|\alpha_v|^2=\pm|\alpha_u|^2$
or $\alpha_u=0.$ As before, the poles with 
$|\alpha_v|^2=+|\alpha_u|^2$ may be omitted as they have pairwise canceling
residues between terms with indices $u$ and $v$ interchanged. The poles
with $|\alpha_v|^2=-|\alpha_u|^2$  correspond to
\eqn\xipole{\xi\ =\ {-(|{\rm v}_u|^2+|{\rm v}_v|^2+2r^2)+
\sqrt{4(r^2-\Re {\rm v}_u\abar_v)^2+|{\rm v}_u^2-{\rm v}_v^2|^2}\over2ir(\abar_u+\abar_v)}
\ =\ {i\over2r}\,({\rm v}_u+{\rm v}_v)+\CO(r^{-3})\ .}
These contribute
\eqn\contribone{\pi^3\sum_{u=1}^N\,\sum_{v\neq u}\,{1\over({\rm v}_v-{\rm v}_u)^2}\,
\prod_{w\neq u,v}\,{\pi^2\over2}
\,{1\over({\rm v}_w-{\rm v}_u)({\rm v}_w-{\rm v}_v)}\,\prod_{f=1}^{N_F}\big(-{1\over\sqrtwo}({\rm v}_u+{\rm v}_v)
+m_f\big)\ +\ \CO(r^{-2})}
to the $\theta$ integral in Eq.~\Fddef. Likewise, the poles at $\alpha_u=0,$
corresponding to $\xi=i{\rm v}_u/r,$ contribute
\eqn\contribtwo{-2\pi\sum_{u=1}^N\,\prod_{v\neq u}\,{\pi^2\over2}
\,{1\over({\rm v}_v-{\rm v}_u)^2}\,\prod_{f=1}^{N_F}(-\sqrtwo {\rm v}_u+m_f)\ +\
\CO(r^{-2})}
to the $\theta$ integral. Adding these two contributions gives, finally:
\eqn\Fddefinal{\eqalign{\calF_\partial\ =\
& -{iC_1'\pi^{2N-1} \over2^{N+2}}
\,\sum_{u=1}^N\,\Big\{\,\sum_{v\neq u}\,
{1\over({\rm v}_v-{\rm v}_u)^2}
\,
\prod_{w\neq u,v}{1\over({\rm v}_w-{\rm v}_u)({\rm v}_w-{\rm v}_v)}
\cr&\times\ 
\prod_{f=1}^{N_F}\big(-{1\over\sqrtwo}({\rm v}_u+{\rm v}_v)
+m_f\big)\ -\ \prod_{v\neq u}\,{1\over({\rm v}_v-{\rm v}_u)^2}\,
\prod_{f=1}^{N_F}(-\sqrtwo {\rm v}_u+m_f)\ \Big\}\ .}}

Despite appearances, it can be shown that this expression vanishes
identically for $N_F<2N-2$. To prove this, it suffices to show that
the residues of all the  simple and double poles cancel among the
various terms,  so that the rational function $\calF_\partial$ must
actually be a polynomial in the variables $\{{\rm v}_u,m_f\}$ (i.e., it must
have a constant denominator). Naive power counting shows that this
polynomial has degree $N_F-2N+2$ so that necessarily
$\calF_\partial\equiv0$ for $N_F<2N-2,$ as stated.

Notice further that the final term in Eq.~\Fddefinal\ precisely cancels
$\calF_\delta$ as given in Eq.~\Fdelfinal. This leaves for the final
one-instanton expression for the prepotential:
\eqn\prepfinal{\eqalign{
\Foneinst\ \equiv\ \calF_\delta+\calF_\partial\ &=\ 
 -{iC_1'\pi^{2N-1} \over2^{N+2}}
\,\sum_{u=1}^N\,\sum_{v\neq u}\,{1\over({\rm v}_v-{\rm v}_u)^2}
\cr&\times\
 \prod_{w\neq u,v}{1\over({\rm v}_w-{\rm v}_u)({\rm v}_w-{\rm v}_v)}\cdot
\prod_{f=1}^{N_F}\big(-{1\over\sqrtwo}({\rm v}_u+{\rm v}_v)
+m_f\big)\ .}}
We reiterate that the product over $N_F$ flavors is to be replaced by
unity when $N_F=0\,$; similarly the product over $w\neq u,v$ is
to be replaced by unity when $N=2$.

As a simple check, notice that for the special
case of $SU(2)$ with $N_F>0$ and all the masses $m_f=0,$ this expression
vanishes identically, since $\vhiggs_1+\vhiggs_2=0$ by the tracelessness
condition. This agrees with the $\bigZ_2$ symmetry arguments in
\refs{\SWtwo,\dkmfour} for this gauge group (see footnote 1); 
the first nonvanishing
contribution in this case is at the 2-instanton level.

\newsec{Discussion of the One-Instanton Result}

\subsec{Comparison with an earlier instanton calculation}

Our final expression for the 1-instanton contribution to the prepotential,
Eq.~\prepfinal, agrees with a previous 1-instanton calculation
by Ito and Sasakura \IStwo\ which used
the 't Hooft-Bernard measure \refs{\tHooft,\Bernard} rather than 
Eq.~\dmutwodef.
 As stated earlier,
these authors made two simplifying assumptions: (1) they assumed that
the final answer depends only 
on the VEVs $\{\vhiggs_1,\cdots,\vhiggs_N\}$ and not on the
complex conjugate parameters $\{\vbarhiggs_1,\cdots,\vbarhiggs_N\}$
(a property known as holomorphy). This allowed them to set the latter
parameters to special tractable values. \hbox{(2) Furthermore,}
  they only extracted
the terms in the integral that become maximally singular in the limit that
 two of the VEVs approach one another. 
 It is a nice property of Eq.~\prepfinal\ for $N_F<2N-2$ and also $N_F=
2N-1$ that this most-singular approximation, when symmetrized in the
VEVs,  reproduces the full rational function.

In the calculation of Sec.~8 above, thanks to the collective coordinate 
measure \dmutwodef,
we were able to drop both these simplifying assumptions. The reason
is the intrinsic simplicity of the (super-)ADHM collective coordinate
parametrization:
the integration variables are all Cartesian,
endowed with a flat measure save for the $\delta$-function insertions.
Consequently, we were able to
 derive, rather than assume, holomorphy.\foot{In the special case
of $SU(2)$, this holomorphy property
is built into the instanton calculus from the
outset: it emerges from a simple rescaling of the bosonic and fermionic
collective coordinates in the $k$-instanton
action \dkmsw. But for $SU(N)$ with $N>2$
no such rescaling removes the $\vbarhiggs_u$ from the problem, and the
ultimate emergence of the purely holomorphic answer  \prepfinal\
seems miraculous from the instanton approach.}

In contrast, for 
$N_F=2N-2$ and $N_F=2N,$ the methods of \IStwo\ are not sufficient for
general $N$
to rule out ``regular'' terms, meaning  terms that are nonsingular for 
all choices of VEVs. By dimensional analysis, these
regular terms can make  the following additive contributions to $\F_1\,$:
\eqna\regterms
$$\eqalignno{N_F=2N-2\,&:\qquad \F_1\ \rightarrow\ \F_1\ +\ C_{2N-2}
\,\Lambda^2\ ,
&\regterms a\cr
N_F=2N\,&:\qquad \F_1\ \rightarrow\ \F_1\ +\ C_{2N}\,e^{-8\pi^2/g^2}_{}\,
\sum_{u=1}^N\,\vhiggs^2_u
&\regterms b\cr}$$
where $C_{2N-2}$ and $C_{2N}$ are numerical constants. Our result from
Sec.~8 can be expressed as
\eqn\Cresult{C_{2N-2}\ =\ C_{2N}\ =\ 0}
when $\F_1$ has the specified form \prepfinal. This agrees with
an explicit integration performed by Ito and Sasakura for the specific case
of $SU(3)$.

\subsec{Comparison with proposed exact solutions for $N_F<2N$}

We can also compare Eq.~\prepfinal\ to the proposed exact hyper-elliptic
curve solutions
of the $SU(N)$ models contained in Refs.~\refs{\KLTY,\AF} for $N_F=0,$ and
in Refs.~\refs{\HO-\MN} for $1\le N_F\le 2N$; the last of these
references is restricted to $N\le3.$ (For $N_F>2N$ the 
$\beta$-function becomes positive and the microscopic model no longer
makes sense as a fundamental theory.) For $N_F<2N$ these curves are
expressed in terms of the 1-instanton factor $\Lambda^{2N-N_F}$ where
$\Lambda$ is the dynamical scale, and a set of quantum
moduli $u_n$ with $2\le n\le N.$ 
The expected gauge-invariant physical
definition of $u_n$ is
\eqn\unequals{u_n\ =\ \langle\Tr\,A^n\rangle\ .}

Extracting the 1-instanton predictions from these curves is a lengthy
exercise, performed in Refs.~\refs{\IStwo,\DKP}; their results are
as follows.
 For $N_F<2N-2$ or $N_F=2N-1,$ Eq.~\prepfinal\ is in
perfect accord with the curves. But for $N_F=2N-2,$ the three curves of
Refs.~\refs{\HO-\MN} give
values of $C_{2N-2}$ which differ from one another, and from the value
$C_{2N-2}=0$ given in Eq.~\Cresult. Of course, the addition of
a constant 
to the prepotential does not affect the low-energy Lagrangian \WilsF\
which depends only on derivatives of $\F.$ But  a constant shift
does affect the quantum modulus $u_2$ whose
\hbox{$k$-instanton}
 component is  proportional to $\F_k$ via Matone's relation, 
Eq.~\matrelis. It follows that, to restore this relation,
 the parameter $u_2$ in
the curves of Refs.~\refs{\HO-\MN} should be linearly shifted by the 
respective
1-instanton factors $C_{2N-2}\,\Lambda^2\,$; only then can $u_2$
 properly be identified with $\langle\Tr\,A^2\rangle$.
A similar shift in $u_2$ must be implemented in the curve of
Seiberg and Witten \SWtwo, at the \hbox{2-instanton level},
 for $N=2$ with $N_F=3$
\AHSW. Likewise $u_3$ must be shifted at the 1-instanton level, for 
$N=3$ with  $N_F=3$ or $N_F=5$,
 in the curves of Refs.~\refs{\HO-\MN},	in order
to restore the identification $u_3=\langle\Tr\,A^3\rangle$ \Slater.

Generically, one should expect such linear shifts in the $u_n$ when
the shifts involve the addition of regular terms, such as shown
in Eq.~\regterms{}
for $u_2$ (recall Eq.~\matrelis).
This implies the following arithmetic. On the one hand, 
from the engineering
dimensions, such quantum
shifts in $u_n$ can only be proportional to
$u_m\,\Lambda^{n-m}$ where $0\le m<n\,$; for an $SU(N)$ rather than a $U(N)$
theory we further require $m\neq1$ since $u_1\equiv0.$ On the other
hand, a $k$-instanton effect is proportional to $\Lambda^{(2N-N_F)k}.$
Consequently, equating powers of $\Lambda,$
we generically expect a $k$-instanton additive shift
to $u_n$ when $k$, $N$ and $N_F$ satisfy 
\eqn\generick{n-m\ =\ (2N-N_F)k\ ,\quad 0\le m<n\ ,\ m\neq1\ .}
Notice that all the explicit examples discovered to date, summarized in the
previous paragraph, fit into this classification.
In contrast, the models with $N_F=2N$ are much more complicated: \it all \rm
instanton orders $k$ can in principle
contribute linear shifts to \it all \rm the $u_n,$
for reasons we now discuss.

\subsec{Comparison with proposed exact solutions for $N_F=2N$}

\def\taumicro{\tau_{\rm micro}^{}}
The models with $N_F=2N$ are finite theories; the
$\beta$-function vanishes, and no dynamical scale is generated. Instead,
 the curves are functions of a dimensionless complexified coupling
$\tau$. Thus the
 dimensional analysis of the previous paragraph
no longer applies; in order to agree with conventional definitions
of condensates $u_n$ and effective couplings $\taueff$,
parameters in the curves can in principle be
shifted at all instanton orders, i.e, by a Taylor series in the
dimensionless 1-instanton parameter $q=\exp(i\pi\taumicro)$, where
$\taumicro$ is the renormalized coupling of the microscopic $SU(N)$
theory (see Ref.~\dkmfive\ for a detailed discussion).

\def\taueffz{\taueff^{\sst(0)}}
\def\tauSW{\tau_{\sst\rm SW}^{}}
The first example of such a  redefinition appears in the
 curve of Seiberg and Witten for $N=2$ with $N_F=4$ \SWtwo.
As shown in Refs.~\refs{\dkmfour,\dkmfive}, the parameter $\tauSW$
that appears in the massive curve, rather than being the microscopic
coupling $\taumicro,$ is actually the effective $U(1)$ coupling evaluated
at the special conformal point in the moduli space where the four bare
hypermultiplet masses vanish.
We termed this effective coupling $\taueffz$ where the superscript reminds us
of this masslessness condition. 
The relation between these parameters reads \dkmfive:
\eqn\taurels{\tauSW\ =\ \taueffz\ =\ \taumicro\ +\ {i\over\pi}\,
\sum_{k=0,2,4\cdots}c_k\,q^k\ ,\qquad q=\exp(i\pi\taumicro)\ .}
That the sum runs over only even instanton sectors is due to a
$\bigZ_2$ symmetry specific to the  $SU(2)$ models with $N_F>0$ massless
hypermultiplets \refs{\SWtwo,\dkmfour}. As calculated in \dkmfive,
the contribution to the sum from $k=0$ is
a 1-loop perturbative
effect which comes from a standard application of
Weinberg's matching prescription \RGorig. A formal expression, in quadratures,
for the constants $c_k$ as $k$-instanton
collective coordinate integrals is given
in Ref.~\dkmten; generically, for even $k$, the $c_k$ are all nonzero.
A similar all-even-instantons relation exists
between $\utilde$ (the parameter in
the Seiberg-Witten curve) and $u_2=\langle\Tr\,A^2\rangle$ \dkmfive.
Note that the series \taurels\ in no way contradicts the conformal
invariance of the model, since the right-hand side is a purely
numerical, scale-independent renormalization of the effective coupling
\refs{\dkmfour,\dkmfive}.

Next we discuss the curves for $SU(N)$ gauge theory with $N>2$ and $N_F=2N$. 
Three ostensibly different such curves are proposed in 
Refs.~\refs{\HO-\MN}.\foot{According to Refs.~\refs{\MNII,\AStwo}, 
the curves in \APS\ and \MN\ for $SU(3)$
can be transformed into one
another by a modular redefinition of their respective $\tau$ parameters,
  but no such transformation has yet been found that equates
these curves, in turn, with that of \HO.}
For none of the three curves can the $\tau$ parameters be equated
with $\taumicro.$ This can be seen already at the 1-instanton level:
all three curves give values of $C_{2N}$ different from one another,
and different from the value $C_{2N}=0$ given in Eq.~\Cresult\ \IStwo. 

What, then, is the physical meaning of the $\tau$ parameters in these curves?
By analogy with the $SU(2)$ case \taurels, it is natural to guess
that these $\tau$'s  should be equated to some effective coupling,
$\taueff$, 
rather than to $\taumicro.$ The trouble with such an identification is
that, for $SU(N)$ with $N>2,$ the effective
coupling is an $(N-1)\times(N-1)$ dimensional
 \it matrix \rm rather than a scalar;  furthermore, it is VEV
dependent (equivalently, $u_n$ dependent):
\eqn\taumatrices{(\taueff)^{}_{u,v}\ =\ \hf {\partial^2\over\partial\vhiggs_u
\,\partial\vhiggs_v}\ \F(\{\vhiggs_w\})\ .}
How then should the results of (multi-)instanton calculations enter into these
curve parameters?

A potential answer to this
 question can be given  for the special case
of $SU(3)$. Here there is a distinguished line in  moduli space where the
$2\times2$ 
matrix $\taueff$ is effectively \hbox{1-dimensional}
 \MNII. This is the line where
the six bare  hypermultiplet masses are zero and 
the three VEVs $\vhiggs_{1,2,3}$ are proportional to the cube-roots
of unity (i.e., $u_2=0$ but $u_3\neq0$).
 On this line, the matrix of effective couplings is proportional
to the classical form \MNII\foot{This is what  one gets by
 starting from the classical form of the prepotential which
gives $\tau_{u,v}\propto
\delta_{u,v}$ in the 3-dimensional VEV space, then
imposing on the prepotential
the tracelessness condition $\vhiggs_3=-\vhiggs_1-\vhiggs_2.$
For general $N$, eliminating $\vhiggs_N$ in this way gives $(\taueff)_{u,v}
\propto\delta_{u,v}+1$.}
\eqn\classtau{\taueff\ \propto\ \pmatrix{2&1\cr1&2}}
Reasoning by analogy with the $SU(2)$ case, we believe that a relation
similar to Eq.~\taurels\
 between the  $\tau$ parameters in the curves, and $\taumicro,$
can be obtained by examining this special line in the moduli space.
Similar all-instanton-orders redefinitions of the $u_n$  also
need to be made, as in the $SU(2)$ case \dkmfive.

In contrast, for $N>3$ with $N_F=2N$, it can be proved that there are no points
on the moduli space where $\taueff$ is proportional to the classical
form $\delta_{u,v}+1$ \refs{\MNII,\Yank}.
The authors of \MNII\ argue that the corresponding
curves are actually underdetermined, partially because of too many
available modular forms. 
From the instanton perspective, 
in these cases we do not currently understand 
how to reconcile the $\tau$ parameters in the curves of \refs{\HO,\APS}
with the explicit multi-instanton results, starting with the
1-instanton expression \prepfinal\
above. Furthermore, since the $\tau$ parameters used in Refs.~\refs{\HO,\APS}
do not have an obvious field-theoretic meaning (they are neither microscopic
nor effective couplings as discussed above), we do not understand the
origin of the $\tau\rightarrow-1/\tau$ duality built into these
curves.

Nevertheless, we can offer the following interesting observation, which
might be a clue to the eventual resolution of these issues. Consider the
case of $SU(4)$ with $N_F=8.$ Let us examine the special line in moduli
space where all eight hypermultiplet masses are zero, and the four
VEVs are proportional to the fourth roots of unity (i.e.,
$u_2=u_3=0$ but $u_4\neq0$). At the classical level, 
the matrix
$\taueff$ is proportional to
\eqn\mataa{\pmatrix{2&1&1\cr1&2&1\cr1&1&2}}
by definition (see footnote 14). As shown in \MNII, 
at the 1-loop perturbative 
level, on this line in moduli space, $\taueff$ is corrected
by an amount proportional to the matrix
\eqn\matbb{\pmatrix{0&-2&2\cr-2&-4&-2\cr2&-2&0}}
We have extracted the 1-instanton contribution to $\taueff$ with this
choice of moduli (a simple
calculation using Eqs.~\prepfinal\ and \taumatrices). Intriguingly, 
while this contribution is
not proportional to the classical matrix \mataa, it does turn out to
be proportional to the 1-loop form \matbb. Should this coincidence
persist to arbitrary multi-instanton levels, it
 suggests that
an all-instanton-orders renormalization of the coupling of
the type \taurels\ may
in fact be possible.

Finally we comment on the $M$-theory picture due to Witten \Witten. One might
hope that the construction in \Witten\ provides an unambiguous identification
of the coupling of the theory in terms of the $\tau$ parameter of the
Riemann surface. In fact, if one identifies the asymptotic
separation $\Delta x_6$ between the 4-branes in the type IIA picture with
the microscopic coupling of the theory, one can in turn relate it to the
$\tau$-parameter of the Riemann surfaces associated with the
curves of Refs.~\refs{\HO-\MN}. However, it appears that the naive
identification $\Delta x_6\equiv\taumicro$ cannot be quite right. This
can be most easily seen  in the context of $SU(2)$ gauge theory
with four massless flavors; as shown in Ref.~\ENR, one calculates
\eqn\lisacalc{\Delta x_6\ =\ \tau+\hbox{const.}+\hbox{(instantons)}}
where the first nonvanishing instanton term is at the $k=1$ level.
The parameter $\tau$ here is the parameter in the elliptic curve, which
is known to equal $\taueffz$ \dkmfive\ (see Eq.~\taurels\ above). 
Comparing Eqs.~\taurels\ and \lisacalc, and noting the presence of
1-instanton corrections on the right-hand side of the latter equation,
we conclude that $\Delta x_6$ cannot equal $\taumicro$ precisely;
at best, they can only be equated in the weak-coupling limit.

$$\scriptstyle{**************************************************}$$

We thank A. Hanany, J. Minahan, A. Shapere, and N. Sasakura
 for clarifying discussions
about their work. We are especially
grateful to N. Dorey for numerous enlightening
conversations. 

\vfil\eject
\appendix{A}{Details of Cluster Decomposition of the Measure}

\def\abar{\bar a}
In this appendix we demonstrate the clustering property
of the $U(N)$ $k$-instanton measure. We proceed along the lines of
 \dkmten. 
We will analyze the limit in which one of the instanton position moduli is
far away from all the others, and demand that the measure
factor approximately into a product of a 1-instanton and a
$(k-1)$-instanton measure. Recall that in the limit of large
separation, the space-time positions of the $k$ individual
instantons making up the topological-number-$k$
configuration may simply be identified with the $k$ diagonal elements
$a'_{ii}$ \CWS.
The matrix $a'$ is
understood to be a $k \times k$ matrix with $2 \times 2$ 
quaternion-like entries
$a'_{ij} = (a'_m)_{ij} \ \sigma^m$ (cf.~Eq.~\dec).

Where the unidentified $k$-instanton
measure $d\mu^{(k)}$ is concerned, it is important to understand
cluster decomposition as a $U(k)$ invariant effect.
To achieve this we take the $a'_{kk}$-dependent submatrix of $a'$,
\eqn\hdef{h\ =\ a'_{kk} \cdot
\pmatrix{0\cr\vdots\cr0\cr1}
{\big(0,\cdots,0,1\big)\atop{}}
\ , }
and act on it with the residual $U(k)$ ADHM symmetry
(cf.~Eq. \restw),
\eqn\htf{h \ \to \ g^{\dagger} \ h \ g \ . }
There is a $U(k-1) \times U(1)$ subgroup of $U(k)$ that
leaves $h$ invariant, so that in fact $g$ is restricted to the  
coset $U(k)/\big(U(k-1) \times U(1)\big)$. Choosing
the parametrization
\eqn\notgdef{
g=\exp\pmatrix{0&\cdots&0&-\alpha_{1k}\cr
\vdots&{}&\vdots&\vdots\cr
0&\cdots&0&-\alpha_{k-1,k}\cr
\bar{\alpha}_{1k}&\cdots&\bar{\alpha}_{k-1,k}&0\cr}\ , }
where the $\alpha_{ik}$ are complex numbers,
the action of this coset on $h$ is given by
\eqn\gacts{\eqalign{a'_{kk}\cdot
\pmatrix{0\cr\vdots\cr0\cr1}
\matrix{\big(0,\cdots,0,1\big)\cr{}\cr{}\cr{}}
\ &\longrightarrow\
a'_{kk}\,g^{\dagger}\cdot
\pmatrix{0\cr\vdots\cr0\cr1}
\matrix{\big(0,\cdots,0,1\big)\cr{}\cr{}\cr{}}
\cdot g
\cr
&=\ a'_{kk}\cdot\pmatrix{0&\cdots&0&\alpha_{1k}\cr
\vdots&{}&\vdots&\vdots\cr
0&\cdots&0&\alpha_{k-1,k}\cr
\bar{\alpha}_{1k}&\cdots&\bar{\alpha}_{k-1,k}&1\cr}\ +\ \CO(|\alpha|^2)\ .}}
The second line describes the infinitesimal action of the coset
$U(k)/\big(U(k-1) \times U(1)\big)$ on the matrix $h$.

The transformation \gacts\ enables us to consider the large $|a'_{kk}|$
limit of the unidentified measure in a meaningful way. We can
precisely state the clustering condition as
\eqn\demand{d\mu^{(k)}\ \buildrel{|a'_{kk}|\rightarrow\infty}\over
\longrightarrow\ 
d\mu^{(k-1)}\times d\mu^{(1)}\times dS^{2(k-1)}\ , }
where the volume form $dS^{2(k-1)}$ is just the Haar measure
for the coset $U(k)/\big(U(k-1) \times U(1)\big)$.
We note here the result \gilmore\ that for infinitesimal $\alpha_{ik}$, 
\eqn\infhaar{
dS^{2(k-1)}=\prod_{i=1}^{k-1}d^{2}\alpha_{ik}.
}

In the light of the above analysis, we proceed to set

\def\ahat{{\hat a}}
\eqn\hatadef{a'_{ik}
\ =\ a'_{kk} \, \ahat_{ik}\ 
,\qquad 1\le i\le k-1\ ,}
and to split $\ahat_{ik}$
into a scalar (S) part and a non-scalar (NS) part:
\def\as{\hat a^{\sst\rm S}}
\def\ans{\hat a^{\sst\rm NS}}
\eqn\sdef{\ahat_{ik}\, =\, \as_{ik}+\ans_{ik}\ ,\quad
{\hat a}^{\sst\rm S}_{ik} \,=\, ({\hat a}_0)_{ik} \
\sigma^0 \ ,\quad
{\hat a}^{\sst\rm NS}_{ik} \,=\, \sum_{m=1}^{3} ({\hat a}_m)_{ik} \
\sigma^m \ .}
This change of variables has the effect
\eqn\measurehat{\int
\prod_{i=1}^{k-1}d^8a'_{ik}\ =\ |a'_{kk}|^{8(k-1)}\int\prod_{i=1}^{k-1}
d^6\ans_{ik}\,d^2\as_{ik}\ ,}
and the $\as$ variables are identified with the infinitesimal
group transformation parameters $\alpha$ above. Then by virtue of
Eq. \infhaar\ above, we straightforwardly extract the expected
volume form, $dS^{2(k-1)}$, from the measure.

To further verify \demand, we must now examine the $\delta$-function
constraints in the clustering limit.
We first examine the $\N=1$ measure, given by Eq.~\dmudef.
Ignoring the infinitesimals $\as$, the constraint on purely bosonic
collective coordinates, Eq.~\bcns{a}, is written as
\eqn\bcnsdecmp{\eqalign{
&\prod_{c=1}^3 \ \delta^{(k^2)}\big(\hf \trtwo \tau^c (\abar a) \big)
\cr&=
\prod_{c=1}^3  \prod_{i=1}^{k-1} \delta^{(2)}
\big(\hf \trtwo \tau^c \big( (\wbar w)_{ik} +
\sum_{j=1}^{k-1} \abar'_{ij} \, a'_{kk} \,
\ans_{jk} -
|a'_{kk}|^2 \, \ans_{ik} \big) \big)
\cr&\times \prod_{c=1}^3 \big[\prod_{i=1}^{k-1}
\delta\big( \hf \trtwo \tau^c \big( (\bar{\tilde{a}} \tilde{a})_{ii} -
|a'_{kk}|^2 \, \ans_{ik} \, \ans_{ki} \big) \big)
\big] \big[ \prod_{i < j}^{k-1}
\delta^{(2)} \big( \hf \trtwo \tau^c \big( (\bar{\tilde{a}} \tilde{a})_{ij} -
|a'_{kk}|^2 \, \ans_{ik} \, \ans_{kj} \big) \big) \big]   
\cr&\times \prod_{c=1}^3 \delta\big(\hf \trtwo \tau^c
\big( (\wbar w)_{kk} -
|a'_{kk}|^2 \sum_{j=1}^{k-1} \ans_{kj} \, \ans_{jk} \big)
\big)  \ .
}}
Here $\tilde{a}$ is the matrix left behind when $a$ has
its last row and column removed.
 The \hbox{$\delta$-functions} comprising the
first line on the right hand side of the above equation are used to
saturate the integration over the $\ans$ variables. The effect of this
integration is two-fold. Firstly, it introduces a factor
$|a'_{kk}|^{-12(k-1)}$ into the measure. Secondly, it requires
the replacement of ${\ans}_{ik}$ in the other $\delta$-functions
with an $\CO(1/|a'_{kk}|^2)$ quantity. Consequently,
in the limit $|a'_{kk}| \rightarrow \infty$, the $\delta$-functions
on the second and third lines become just the constraints that
appear in the $(k-1)$-instanton and the $1$-instanton measure
respectively.

Turning to the second, Grassmannian, constraint in the $\N=1$
measure, we see that it similarly factorizes into three pieces:
\eqn\fcnstrdecmp{\eqalign{
&\delta^{(2k^2)} \big(\Mbar a + \abar \M \big)
\cr&= 
\prod_{i=1}^{k-1} \delta^{(4)}\big((\mubar w + \wbar \mu)_{ik}
+ \sum_{j=1}^{k-1} ( \bar a'_{ij} \, \M'_{jk} +
\bar\M'_{ij} \, a'_{kk} \, \ans_{jk}) -
\ans_{ik} \, a'_{kk} \, \M'_{kk} + \bar\M'_{ik} \, a'_{kk} \big)
\cr&\times \big[ \prod_{i=1}^{k-1} 
\delta^{(2)} \big( \sum(\bar{\tilde{\M}} \tilde{a} + \bar{\tilde{a}}
\tilde{\M})_{ii} + \ldots \big) \big]
\big[ \prod_{i < j}^{k-1} \delta^{(4)} \big(
(\bar{\tilde{\M}} \tilde{a} + \bar{\tilde{a}} \tilde{\M})_{ij}
+ \ldots \big) \big]
\cr&\times \ \delta^{(2)}\big( (\mubar w + \wbar \mu)_{kk} + \ldots \big) 
\ .
}}
Here $\tilde{\M}$ is the matrix left behind when $\M$ has
its last row and column removed. The first $\delta$-function
factor above saturates the integration over the Grassmannian collective
coordinates $\M'_{ik} \ (i=1,\ldots,k-1)$. In performing this
integration, a factor $|a'_{kk}|^{4(k-1)}$ is introduced into the
measure. This exactly cancels the factors that appeared earlier.
Further, in the large $|a'_{kk}|$ limit the omitted terms in the arguments of
the second and third $\delta$-function factors in Eq.~\fcnstrdecmp\
vanish, and we are left with precisely the Grassmannian constraints that
appear in $d\mu^{(k-1)}$ and $d\mu^{(1)}$ respectively.
Since the numerical prefactor $C_{1}^k$ will
also factorize correctly, this completes the proof
of the clustering property, Eq.~\demand, for the $\N=1$
$k$-instanton measure.

In the case of the $\N=2$ measure, Eq.~\dmutwodef, there are two further
$\delta$-function constraints to consider. The $\delta$-function constraint
on the Higgsino collective coordinate can be factorized in exactly the same way
as the gaugino constraint, Eq.~\fcnstrdecmp. Then integration over the
$\N'_{ik} \ (i=1,\ldots,k-1)$ yields a Jacobian factor $|a'_{kk}|^{4(k-1)}$
and leaves, in the large $|a'_{kk}|$ limit, the required $1$-instanton
and $(k-1)$-instanton constraints. As for the constraint on $\Atot$, we
can write:
\eqn\Atotdecmp{\eqalign{
&\delta^{(k^2)}\big( \bigL\cdot\Atot - \Lambda - \Lambda_f \big)
\cr&=\delta^{(2(k-1))}\big( |a'_{kk}|^2 \Atot + \ldots \big)
\cr&\times \delta^{((k-1)^2)}\big( \tilde{\bigL}\cdot\tilde{\A}_{\rm tot}
- \tilde{\Lambda} - \tilde{\Lambda}_f +\ldots \big)
\cr&\times \delta^{(1)}\big( \trtwo(\wbar w)_{kk} (\Atot)_{kk} - \Lambda_{kk}
- (\Lambda_f)_{kk} + \ldots \big)
\ , } }
where the tilde represents 
quantities constructed out of the truncated
matrices
$\tilde{a},\ \tilde{\M}, \ \tilde{\N}$ in the obvious manner.
The omitted terms are subleading
in $|a'_{kk}|$. It is therefore clear that after integrating over
$(\Atot)_{ik} \ (i=1,\ldots,k-1)$, we get a Jacobian factor 
$|a'_{kk}|^{-4(k-1)}$
which cancels the previous factor, and the required $1$-instanton and
$(k-1)$-instanton constraints follow.

\listrefs

\bye